\documentclass[prl,twocolumn,notitlepage,superscriptaddress]{revtex4-2}
\usepackage{graphicx,bm,amsmath}
\usepackage[usenames,dvipsnames]{xcolor}
\usepackage[colorlinks,bookmarks=false,citecolor=blue,linkcolor=Red,urlcolor=blue,
]{hyperref}
\usepackage[percent]{overpic}
\usepackage{bbm}
\usepackage{slashed}
\usepackage[bbgreekl]{mathbbol}
\usepackage{xcolor}
\usepackage[normalem]{ulem}
\usepackage{braket}
\usepackage[ruled,vlined]{algorithm2e}
\usepackage{comment}
\usepackage{amsthm}
\usepackage{lipsum}
\usepackage{enumerate}
\usepackage{enumitem}
\usepackage{subfigure}
\graphicspath{{./figs/}}

\newcommand{\newsec}[1]{\textcolor{blue}{{\textit{#1}.-}}}

\newcommand{\Tr}{\text{Tr}}  
\newcommand{\ii}{\mathrm{i}}
\newcommand{\dd}{\mathrm{d}}

\newcommand{\vk}{\mathbf{k}}
\newcommand{\vp}{\mathbf{p}}
\newcommand{\vq}{\mathbf{q}}
\begin{document}
\title{Cavity-Induced Excitonic Insulation and Non-Fermi-Liquid Behavior in Dirac Materials}

\author{Yuxuan Guo}
\email{yuxguo2024@g.ecc.u-tokyo.ac.jp}
\affiliation{Department of Physics, The University of Tokyo, 7-3-1 Hongo, Bunkyo-ku, Tokyo, 113-0033, Japan}
\author{Yuto Ashida}%
\affiliation{Department of Physics, The University of Tokyo, 7-3-1 Hongo, Bunkyo-ku, Tokyo, 113-0033, Japan}
\affiliation{Institute for Physics of Intelligence, The University of Tokyo, 7-3-1 Hongo, Tokyo 113-0033, Japan}

\begin{abstract}
We investigate two-dimensional Dirac fermions embedded in a deep-subwavelength cavity formed by high-impedance metasurfaces. We point out that, unlike conventional metallic boundaries, these metasurfaces support quasielectrostatic transverse-magnetic modes that mediate a long-range interaction between two-dimensional electrons. 
Combining static electronic screening with a Dyson-Schwinger analysis, we show that this engineered interaction can qualitatively alter the ground-state properties of Dirac materials. For a fermion flavor number $N_{f}$ below a critical value $N_{c}=16/\pi$, the interaction drives an excitonic insulating phase through an infinite-order quantum phase transition and spontaneously generates a mass gap. At $N_{f}>N_{c}$, the system remains gapless but enters a non-Fermi-liquid critical regime where the quasiparticle residue is singularly suppressed to zero, and the Dirac cone exhibits a nonanalytic dispersion relation. Furthermore, under a perpendicular magnetic field, the cavity fluctuations dynamically lift the zeroth Landau level degeneracy across all $N_{f}$. These results identify high-impedance metasurface cavities as promising platforms for engineering correlated Dirac matter.
\end{abstract}

\maketitle

\newsec{Introduction}
Electromagnetic fields have long served as a means of controlling matter, usually through external driving. A complementary question is whether or not the vacuum field of a structured electromagnetic environment can itself modify a material ground state~\cite{garcia-vidal2021Manipulating,scheucher2020Cavity,kennes2022NewEra, ruggenthaler2018,flick2019,mueller2020Deep,owens2022,ashida2022Nonperturbative,ebbesen2016,feist2018,ribeiro2018,galego2015,PhysRevX.7.011030}. Since Purcell's observation that radiative decay depends on the photonic environment~\cite{purcell1946Spontaneous}, the idea of cavity quantum electrodynamics (cQED) has now developed from few-level emitters toward many-body settings, including magnetic~\cite{vinasbostrom2023RuCl3,chiocchetta2021Cavityinduced,PhysRevB.110.035158}, superconducting~\cite{curtis2019Cavity,sentef2018Cavity,schlawin2019CavityMediated,curtis2019Cavity,curtis2022Cavity,li2022}, ferroelectric~\cite{ashida2020Quantum,latini2021Ferroelectric}, and topological~\cite{wang2019,rokaj2019Quantum,*rokaj2022Polaritonic,downing2019Topological,tokatly2021,MK22,hubener2021Engineering,DO21,GLA25} materials. 
These developments have raised the possibility that the quantum vacuum may serve not merely as a passive electromagnetic environment, but as a platform capable of controlling phases of matter.

A central question in this area is how engineered vacuum fluctuations can mediate effective interactions strong enough to qualitatively reconstruct a many-body ground state, rather than merely shift spectra or transport coefficients~\cite{bloch2022Strongly,schlawin2022Cavity,hubener2024Quantum,riolo2025Tuning,eckhardt2025Surface,liu2025Cavity,YA23,KB24,NG25}. While ultrastrong light-matter coupling \cite{gambino2014Exploring,chikkaraddy2016Singlemolecule,yoshihara2017Superconducting,bayer2017Terahertz,halbhuber2020Nonadiabatic,genco2018Bright,flick2017Atoms,forn-diaz2019Ultrastrong,friskkockum2019Ultrastrong} is desirable, the enhanced local electromagnetic density of states due to the Purcell effect~\cite{forn-diaz2019Ultrastrong,friskkockum2019Ultrastrong,PhysRevLett.131.013602} does not by itself imply arbitrarily strong coupling to transverse propagating photons. This has motivated the search for deep-subwavelength cavities below the diffraction limit \cite{novotny2012principles, basov2021polariton, reserbat2021quantum}, where the relevant low-energy excitations are
 often evanescent and quasielectrostatic modes rather than propagating photons~\cite{rivera2020Light,riolo2025Tuning,eckhardt2025Surface,helmrich2024Cavity,andolina2025Electrostatics,cheng2025Anyonic}.

In two metallic mirrors separated by a deeply subwavelength distance, however, the low-energy mode is usually a gap-plasmon-like excitation associated with fluctuating surface charges. In the image-charge picture, a charge fluctuation on one mirror induces an opposite charge on the other, producing a zero-point electric field mainly polarized perpendicular to the cavity plane~\cite{cortese2021excitons, Cortese:2018mfc}. Such polarization couples inefficiently to the in-plane motion of  two-dimensional electrons. Thus, the relevant design problem is not only to tightly confine the vacuum field, but also to engineer its polarization and spatial profile.

\begin{figure}[t!]
    \centering
    \includegraphics[width=0.98\linewidth]{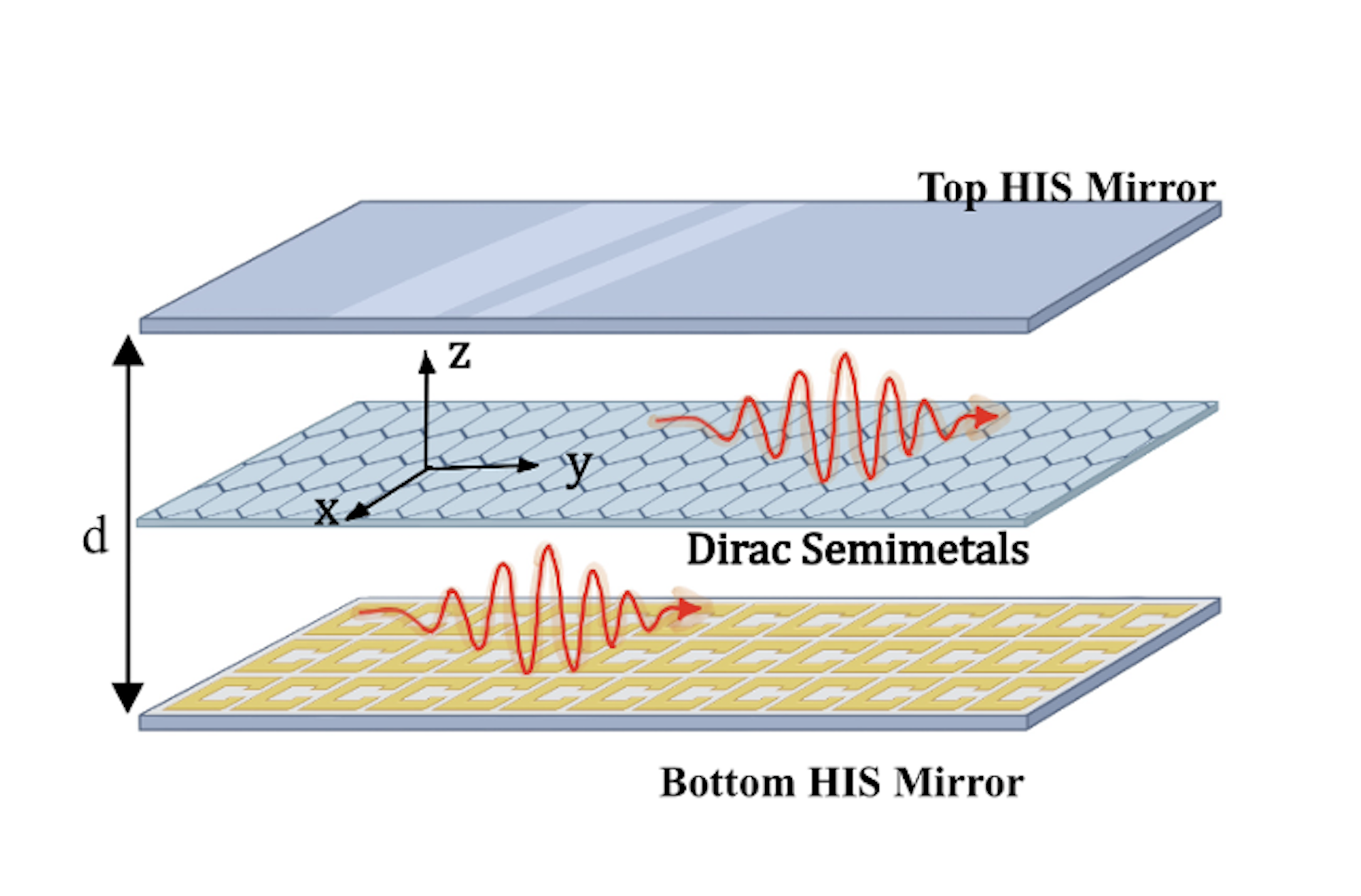}
\caption{Dirac material embedded in a high-impedance metasurface cavity. A quasielectrostatic TM mode mediates a long-range interaction $\mathcal{V}(r)\sim \alpha \ln r$ between electrons at low frequencies near the center $z=0$.  }
\label{fig:overview}
\end{figure}

High-impedance metasurfaces (HISs) offer an ideal route to such control. Over a finite frequency window, they can emulate magnetic-conductor boundary conditions, for which the image charges on the two sides have the same sign~\cite{sievenpiper1999High,sievenpiper2005High,cui2022Spoof}. 
The resulting transverse-magnetic (TM) mode has a strong in-plane electric field near the cavity center and can be described, in the deep-subwavelength regime, as a nearly dispersionless quasielectrostatic mode. Because such metasurfaces are built from subwavelength resonant elements, often describable as effective $\mathrm{LC}$ resonators, their effective impedance and resonance frequency can be tuned without relying on a conventional  Fabry--Perot geometry; representative platforms include split-ring resonators \cite{yen2004terahertz,scalari2012ultrastrong,scalari2014Superconducting,PhysRevLett.105.247402}, graphene plasmonic resonators \cite{fang2014active,ju2011graphene}, and phonon-polaritonic resonators in hexagonal boron nitride \cite{caldwell2014sub-diffractional,dai2014tunable}.

In this work, we consider a two-dimensional Dirac semimetal \cite{castroneto2009graphene, armitage2018weyl, lv2021experimental}, such as graphene, placed in a deeply subwavelength cavity formed by two HISs. The cavity supports a nearly flat quasielectrostatic TM mode that mediates an electron-electron interaction with the potential $\mathcal{V}(r)\sim \alpha \ln r$, in sharp contrast to the ordinary Coulomb form.  
We show that this engineered interaction does not merely renormalize spectra and transport, but can qualitatively alter the many-body ground state of Dirac materials by reshaping the interaction kernel. 
Including electronic screening effect and solving the corresponding Dyson--Schwinger problem, we find that, for a fermion flavor number $N_f<N_c$ with $N_c=16/\pi$, the cavity confinement induces an excitonic mass generation \cite{RevModPhys.84.1067,Khveshchenko:2008ye,PhysRevLett.87.246802} through the spontaneous condensation of particle-hole pairs. In contrast, when $N_f>N_c$, the mass vanishes, and the gapless state exhibits non-Fermi-liquid behavior with no well-defined quasiparticles \cite{Gonzalez:1993uz,PhysRevB.76.115434,altshuler1994low,kim1994gauge,abanov2003quantum,chubukov2003nonanalytic,metzner2003soft,lee2009low}. Specifically, the bare Dirac cone is  reshaped into a nonlinear low-energy dispersion, and the quasiparticle residue is singularly suppressed to zero, reflecting a vanishing overlap with the noninteracting electrons and the loss of a quasiparticle pole. Furthermore, under magnetic field \cite{smolka2014Cavity,zhang2016Collective,paravicini-bagliani2019Magnetotransport,keller2020,appugliese2022Breakdown,ravets2018,hagenmuller2010Ultrastrong,69jg-sq8r}, the same cavity modes lift the degeneracy of the zeroth Landau level (ZLL) at any value of $N_f$, leading to a bulk incompressible state. Taken together, these results identify high-impedance metasurface cavities as promising platforms to realize correlated phases of Dirac materials.

\newsec{High-impedance-surface cavity-mediated interaction}
We consider a planar cavity formed by two identical HISs located at $z=\pm d/2$, with a dielectric spacer of relative permittivity $\varepsilon_d$ in between. 
The cavity is translationally invariant in the in-plane coordinate ${\bm r}=(x,y)$, and a two-dimensional Dirac material is placed inside the cavity, where it couples to the in-plane electric field of the cavity mode. We focus on the $H_y$ antisymmetric TM  branch carrying continuous in-plane momentum $\bm q$, whose field profile maximizes the in-plane coupling to a two-dimensional layer placed at the cavity center.

In the frequency range of interest, the metasurface period $a$ is assumed to satisfy $a\ll \lambda_0$, where $\lambda_0$ is the free-space wavelength. 
Under this condition, the structured mirrors can be described by an effective local surface impedance $Z_s(\omega)$ \cite{sievenpiper1999High} without considering their microscopic subwavelength diffraction pattern. 
The HIS behaves as a reactive surface with an LC-type impedance, which we parametrize as
$Z_s(\omega)=\left(\ii\omega C_s+\frac{1}{\ii\omega L_s}\right)^{-1}$. 
Inside the dielectric spacer, the TM mode with momentum $\vq=q\mathbf{e}_x$ has field components ${\bf E}(\bm r,z,t)=(E_x(z),0,E_z(z))^{\rm T}e^{-\ii{\bf q}\cdot{\bm r}+\ii\omega t}$ and ${\bf H}(\bm r,z,t)=(0,H_y(z),0)^{\rm T}e^{-\ii{\bf q}\cdot{\bm r}+\ii\omega t}$. 
For the antisymmetric TM mode, the impedance boundary condition at $z=\pm d/2$ is $E_x(\pm d/2)=\pm Z_s(\omega)H_y(\pm d/2)$, which yields the exact dispersion equation
\begin{equation}
\coth\!\left(\frac{\kappa_d d}{2}\right)
=
\frac{\omega\varepsilon_0\varepsilon_d\, Z_s(\omega)}{\ii\kappa_d},
\label{eq:dispersion}
\end{equation}
where $\kappa_d^2=q^2-\varepsilon_d\omega^2/c^2$. 
We are interested in the narrow-gap deep-subwavelength regime, where $q\gg \omega/c$ and $qd\ll 1$. 
For a resonant HIS described by the LC surface impedance above, Eq.~\eqref{eq:dispersion} then yields a nearly constant mode frequency,
\begin{equation}
\omega_{\bf q}^2 \simeq \omega_p^2
=
\frac{\omega_0^2}{1+\varepsilon_0\varepsilon_d d/(2C_s)},
\label{eq:omegap}
\end{equation}
where $\omega_0=1/\sqrt{L_sC_s}$ is the bare resonance frequency of an isolated HIS. 
Thus, in the regime of interest, the cavity supports a quasielectrostatic TM mode with the nearly constant frequency at $\omega_p$.

Because the mode is quasielectrostatic, it is convenient to introduce a scalar potential operator $\hat\phi$ such that $\hat{\bf E}\simeq -\nabla\hat\phi$. 
For the antisymmetric cavity mode, the potential profile is proportional to $\cosh(qz)$, and we can write
\begin{equation}
\hat{\phi}({\bm r},z)
=
\sum_{\bf q}
{\cal N}_{\bf q}^{-1/2}\,\cosh(qz)\,e^{-\ii{\bf q}\cdot{\bm r}}
(\hat a_{\bf q}+\hat a^\dagger_{-{\bf q}}),
\label{eq:phiop}
\end{equation}
where the bosonic operators satisfy $[\hat a_{\bf q},\hat a_{{\bf q}'}^\dagger]=\delta_{{\bf q},{\bf q}'}$, and ${\cal N}_{\bf q}=Sq^2(C_s+\epsilon_0\epsilon_d d/2)/(\hbar\omega_{\bf q})$ is the normalization factor with
 $S$ being the in-plane quantization area. 
Here, we quantize the mode by equating its electromagnetic and surface energy to $\hat{H}_{\rm cav}=\sum_{\bf q}\hbar\omega_{p}(\hat a_{\bf q}^\dagger\hat a_{\bf q}+1/2)$, and  $\mathcal{N}_{\bf q}$ is fixed by the total energy per mode, including both the field energy stored in the dielectric slab and the energy stored on the two high-impedance boundaries (see Supplemental Material (SM) \cite{supplemental} for details). 
 
\begin{figure}[t!]
    \centering
  \includegraphics[width=\linewidth]{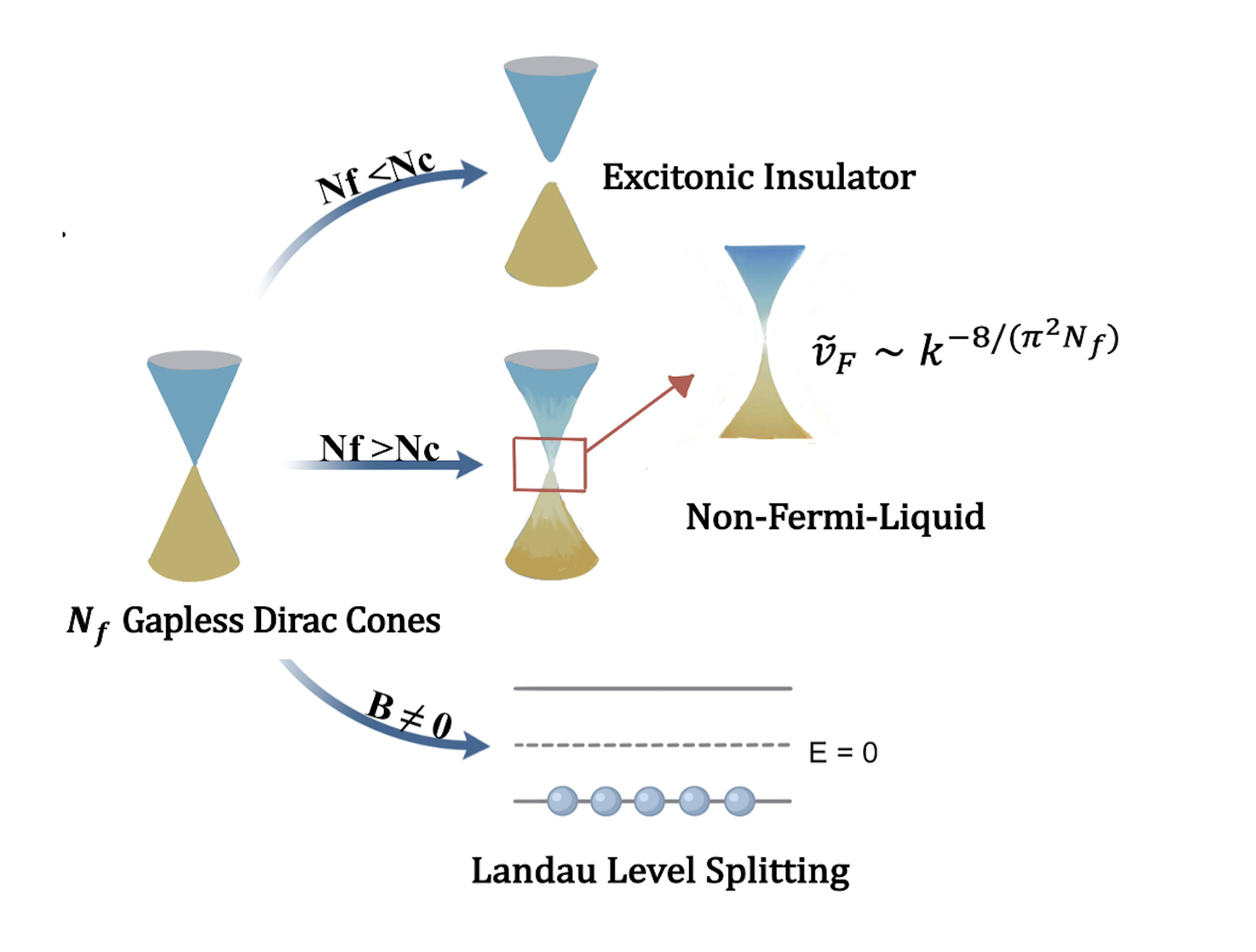}
\caption{Phase landscape realized by the cavity confinement: $N_f < N_c$ marks an excitonic insulator, while $N_f > N_c$ signifies a gapless non-Fermi liquid state where interaction-induced fluctuations make the dispersion relation nonlinear and suppress the quasiparticle. At a nonzero magnetic field $B$, the interaction induces a splitting of the zeroth Landau level at $E=0$  for any fermion flavor number $N_f$. }
\label{fig:phases}
\end{figure}

We now consider placing a two-dimensional Dirac material at the cavity center $z=0$. 
Using a two-component notation, the free-fermion Hamiltonian is written as
\begin{gather}
\hat H_0
=
-\sum_{a=1}^{N_f}\sum_{\bf k}
\hat\Psi^\dagger_{a{\bf k}}\,
v_F\bigl[(-1)^a\sigma_x k_x+\sigma_y k_y\bigr]\,
\hat\Psi_{a{\bf k}},
\label{eq:H0_twocomp}
\end{gather}
where $\hat\Psi_{a,{\bf k}}=(\hat c_{A,a,{\bf k}},\hat c_{B,a,{\bf k}})^{\rm T}$, $A$ and $B$ denote the two sublattices or other internal degrees of freedoms, $a$ sums over other external freedoms (i.e., valley and spin), and $(-1)^a=\pm 1$ labels the chirality of fermions. The Green's function is given by
$
G_0(\ii\omega,{\bf k})
=
(\ii\omega+v_F\boldsymbol{\sigma}\!\cdot{\bf k})
/(\omega^2+v_F^2k^2)
$, where $\boldsymbol{\sigma}=((-1)^a\sigma_x,\sigma_y)^{\rm T}$. Since the static Coulomb interaction is strongly screened by metallic ground plane. We focus on the cavity-electron coupling through
\begin{equation}
\hat H_{\rm int}
=
-\frac{e}{\sqrt S}
\sum_{a=1}^{N_f}
\sum_{{\bf k},{\bf q}}
\hat\Psi^\dagger_{a,{\bf k}+{\bf q}}
\hat\Psi_{a{\bf k}}\,
\hat\phi_{\bf q},
\label{eq:Hint_twocomp}
\end{equation}
where $\hat\phi_{\bf q}\equiv \hat\phi_{\bf q}(z=0)$.

Equations \eqref{eq:phiop}, \eqref{eq:H0_twocomp}, and \eqref{eq:Hint_twocomp} provide the complete low-energy description $\hat {H}_{\rm cQED}=\hat{H}_{\rm cav}+\hat{H}_0+\hat{H}_{\rm int}$ to analyze the ground-state properties. 
At the cavity center, the scalar-potential propagator is
\begin{equation}
D_\phi(\ii\omega,{\bf q})
=
\dfrac{2\omega^2_{p}}{\omega^2+\omega_{p}^2}\,
\dfrac{1}{q^2\left(C_s+\varepsilon_0\varepsilon_d d/2\right)},
\end{equation}
and the exchange of cavity photon mediates the bare interaction
${\mathcal{V}}_0(\ii\omega,{\bf q})=-e^2D_\phi(\ii\omega,{\bf q})$.
We note that, in the low-energy regime $|\omega|\ll \omega_p$, this reproduces the quasistatic form
${\mathcal{V}}_0({\bf q})\simeq -\alpha/q^2$ with an energy scale $\alpha={4e^2}/({2C_s+\varepsilon_0\varepsilon_d d})$.

\newsec{Excitonic insulator phase for $N_f<N_c$}
We now ask how the cavity-induced interaction can destabilize the Dirac semimetal by solving the Dyson--Schwinger equation \cite{PhysRevD.33.3704,Maris:2003vk,Khveshchenko:2008ye,PhysRevLett.87.246802,Appelquist:1988sr,PhysRevB.66.045108} for the dressed Green's function, 
$G^{-1}(\ii\omega,{\bf k})=G_0^{-1}(\ii\omega,{\bf k})-\Sigma(\ii\omega,{\bf k})$, leading to
\begin{equation}
\Sigma(\ii\omega,{\bf k})
\!=\!
-\!\int\frac{\dd\Omega}{2\pi}
\int\frac{\dd^2q}{(2\pi)^2}\,
{\mathcal V}_{\rm RPA}(\ii\Omega,{\bf q})\,
G(\ii\omega+\ii\Omega,{\bf k}+{\bf q}),
\label{eq:Dyson_noncovariant}
\end{equation}
where we use the RPA-screened interaction, ${\mathcal{V}}_{\rm RPA}(\ii\Omega,{\bf q})={\mathcal{V}}_0(\ii\Omega,{\bf q})/[1-\Pi(\ii\Omega,{\bf q}){\mathcal{V}}_0(\ii\Omega,{\bf q})]$ with $\Pi(\ii\Omega,q)$ being the vacuum polarization \cite{supplemental}. 
The most effective suppression of interactions arises from static screening, governed by the static polarization $\Pi(0,{\bf q})=-N_f q/(16v_F)$, which is larger than its finite-frequency counterpart $\Pi(\ii\omega,{\bf q})$.
One may then use the instantaneous approximation \cite{Khveshchenko:2008ye,PhysRevLett.87.246802,PhysRevB.88.205429,PhysRevB.66.045108,Adler:1984ri,alkofer1988chiral}, in which  ${\mathcal{V}}_{\rm RPA}(\ii\Omega,{\bf q})$ is replaced by ${\mathcal{V}}_{\rm RPA}(0,{\bf q})$.  Since this approximation generally underestimates the interaction strength, an instability found within it, if present, may be viewed as robust. 
Substituting the bare cavity-induced interaction ${\mathcal{V}}_0({\bf q})\simeq -\alpha/q^2$, we have
$
 {\mathcal{V}}_{\rm RPA}(0,{\bf q})
\sim 16v_F/(N_fq),
$
which indicates that the long-range interaction is screened to a $1/q$ repulsive interaction at low energies, where $\alpha$ provides the natural UV scale. 

To examine a possible instability, we consider the excitonic channel, characterized by the condensation of particle-hole pairs. The corresponding order parameter is the fermion bilinear $\langle \hat\Psi^\dagger \sigma_z \hat\Psi\rangle$, and we therefore approximate the self-energy as $\Sigma(\ii\omega,{\bf k})\simeq M(k)\sigma_z$. In the low-energy regime,  Eq.~\eqref{eq:Dyson_noncovariant} leads to the nonlinear gap equation that can be brought to a differential form~\cite{supplemental}.
\begin{equation}
\frac{\dd}{\dd k}
\left(
k^2 \frac{\dd M(k)}{\dd k}
\right)
+
\frac{4v_Fk}{\pi N_f}\,\frac{M(k)}{\sqrt{M^2(k)+v_F^2k^2}}
=
0,
\label{eq:gap_diff_eq}
\end{equation}
with boundary conditions
$
k^2 d_{k}M|_{v_Fk=M(0)}=0
$ and
$
\left[kd_k M+M\right]_{v_Fk=\alpha}=0.
$
Here the infrared cutoff is set self-consistently by the excitonic gap itself, while the ultraviolet scale $\alpha$ marks the breakdown of the low-energy $1/q$ kernel inherited from the screened cavity interaction. 
Within the present convention, the critical flavor number is $N_c=16/\pi$; when  $N_f<N_c$, Eq.~\eqref{eq:gap_diff_eq} admits oscillatory solutions that can be matched to both boundary conditions, yielding that the condensation of particle-hole bound-state drives the system into an excitonic insulator phase with nonzero gap. 
The resulting energy gap obeys 
\begin{equation}
M(0)=
\alpha\,
\exp\!\left[
-\frac{2\pi-\delta_{N_f}}{\sqrt{N_c/N_f-1}}
\right],
\label{eq:M0_scaling}
\end{equation}
where $\delta_{N_f}=2\arctan(\sqrt{N_c/N_f-1})$. In contrast to previous studies, such as Refs.~\cite{liu2025Cavity, wang2019Cavity,cheng2025Anyonic}, where gap opening typically relies on explicit symmetry breaking imposed by the photonic environment itself, the mass gap in our theory originates from the spontaneous breaking of the inversion symmetry corresponding to $\bf r\rightarrow-\bf r$ and the exchange of $A$ and $B$.

\newsec{Gapless critical phase for $N_f>N_c$}
When $N_f>N_c$, the gap equation admits only the trivial solution $M(k)=0$, and the Dirac spectrum remains gapless. This does not, however, imply an ordinary Dirac semimetal. While the cavity-mediated interaction is insufficient to generate a mass in this regime, it still produces substantial self-energy corrections leading to a critical phase with  non-Fermi-liquid properties.

To characterize this regime, we parameterize the dressed propagator as
$
G(\ii\omega,{\bf k})
=\tilde{Z}/(-\ii\omega
+\tilde v_F{\boldsymbol \sigma}\cdot{\bf k})
$ and $\Sigma(\ii\omega,{\bf k})
=
\tilde{Z}^{-1}\left[\ii\omega\,\Sigma_\omega
+
v_F\,\boldsymbol{\sigma}\!\cdot\!{\bf k}\,\Sigma_k\right]$, where $\tilde{Z}$ and $\tilde{v}_F$ are the running quasiparticle residue and Fermi velocity, respectively. To evaluate the one-loop fermion self-energy, we conduct a self-consistent RPA approximation \cite{son2007quantum, dassarma2007theory, foster2008graphene, gonzalez1999marginal, kotov2008electron, polini2008plasmons}, where $\tilde{\Pi}(\ii\omega,\vq)=\frac{\tilde{Z}^2N_fq^2}{\sqrt{\tilde{v}_F^2q^2+\omega^2}}$ and substitute it into Eq.~\eqref{eq:Dyson_noncovariant}. In the infrared regime, the frequency-dependent part reduces to
\begin{equation}
\Sigma_\omega(\mu)
=
-\frac{8}{\pi^2N_f }
\int_{\mu/v_F}^{\alpha/v_F}\frac{\dd q}{q}
\int_0^{\alpha/(v_Fq)}\dd x\,
\frac{1-x^2}{(1+x^2)^{3/2}},
\label{eq:Sigmaomega_xq_revised}
\end{equation}
where $x=\Omega/(v_F q)$ and $\mu$ is the running energy scale. The spatial part has the analogous form
\begin{equation}
\Sigma_k(\mu)
=
-\frac{8}{\pi^2N_f }
\int_{\mu/v_F}^{\alpha/v_F }\frac{\dd q}{q}
\int_0^{\alpha/(v_Fq)}\dd x\,
\frac{x^2}{(1+x^2)^{3/2}}.
\label{eq:Sigmak_xq_revised}
\end{equation}
The dressed quasiparticle residue and Fermi velocity are then
$
\tilde {Z}(\mu)=({1+\Sigma_\omega(\mu)})^{-1}$ and 
${\tilde v_F(\mu)}=
{v}_F\frac{{1-\Sigma_k(\mu)}}{{1+\Sigma_\omega(\mu)}}$, respectively. Importantly, the combination of $\Sigma_k(\mu)$ and $\Sigma_\omega(\mu)$ is obtained at the leading order in $1/N_f$ without any ambiguity from gauge fixing, where
$
\Sigma_k(\mu)+\Sigma_\omega(\mu)
=
-\frac{8 }{\pi^2N_f}\ln\frac{\alpha}{\mu}.
$
This implies the renormalization group (RG) flow of the Fermi velocity
\begin{equation}
\frac{\dd \ln \tilde v_F}{\dd \ell}
=
\frac{8}{\pi^2N_f}
+
O\!\left(\frac{1}{N_f^2}\right),
\end{equation}
where $\ell=\ln\frac{\alpha}{\mu}$ is the logarithmic RG scale. The low-energy dispersion therefore departs from the bare linear Dirac form through a singular enhancement of the velocity as
\begin{gather}
    \tilde{\omega}(k)\propto v_F\ k^{1-\,8/(\pi^2N_f)}.
\end{gather}

The quasiparticle residue is also suppressed by the interaction as
$
\tilde {Z}(\omega)\propto
{(\omega/\alpha)^{\frac{4\ln (\omega/\alpha)}{\pi^2 N_f}}}.
$
Thus, while no true gap opens for $N_f>N_c$, the singular velocity makes the single-particle Green's function lose its well-defined quasiparticle pole, and the quasiparticle coherence is progressively reduced to zero at low energies. Namely, the system enters a non-Fermi-liquid critical regime where the sharp Dirac spectral peak is washed out and replaced by an incoherent low-energy continuum. Technical details of the one-loop self-energy calculation and the associated RG analysis can be found in the SM \cite{supplemental}.

\newsec{Interaction-induced splitting of the zeroth Landau level}
Finally, we consider the Dirac system subject to a strong perpendicular magnetic field, which reorganizes the low-energy spectrum into discrete Landau levels. The low-energy Hamiltonian with $(-1)^a=1$ is
$
\hat H_{B,+}
=
v_F
\begin{pmatrix}
0 & \hat{\Pi}_x-\ii\hat{\Pi}_y\\
\hat{\Pi}_x+\ii\hat{\Pi}_y & 0
\end{pmatrix},
$
where $\bm{\hat{\Pi}}=\bm{\hat{p}}+e\bm{\hat{A}}$ is the kinetic momentum; similarly, the Hamiltonian with $(-1)^a=-1$ is given by $\hat{H}_{B,-}=\hat{H}_{B,+}^T$.
Introducing the ladder operators
$
\hat a=\frac{\ell_B}{\sqrt{2}}(\hat\Pi_x-\ii\hat\Pi_y),
\quad
\hat a^\dagger=\frac{\ell_B}{\sqrt{2}}(\hat\Pi_x+\ii\hat\Pi_y),
$
with $\ell_B=\sqrt{\hbar/eB}$, one obtains the corresponding Landau level spectrum
$
E_{n,\pm}=\mathrm{sgn}(n)\sqrt{|n|}\omega_B$, where $\omega_B=\,v_F\sqrt{2}/\ell_B.
$
The ZLL lying exactly at zero energy is macroscopically degenerate and polarizes onto a single sublattice component for a fixed chirality, making it particularly susceptible to interaction effects.

We consider how the cavity-induced interaction can lift this degeneracy and alter the low-energy physics. The low-energy dynamics is dominated by the ZLL, whose  Green's function factorizes into a gauge-dependent Schwinger phase and a translationally invariant part \cite{miransky2015quantum},
$
G_{\rm ZLL}(\ii\omega;{\bf r},{\bf r}')
=
e^{\ii\Phi({\bf r},{\bf r}')}
\widetilde G_{\rm ZLL}(\ii\omega;{\bf r}-{\bf r}').
$ We note that the noninteracting projected propagator is massless:
$
\widetilde G_{0,\rm ZLL}(\ii\omega,{\bf q})
=
2\,e^{-q^2\ell_B^2}\,
\frac{1}{-\ii\omega}\,
\mathcal P_\pm,
$
where $\mathcal P_\pm=\frac{1\mp\sigma_z}{2}$ projects onto the sublattice component supporting the ZLL.
The corresponding real-space Dyson-Schwinger equation for the interaction-dressed ZLL propagator is \cite{Gusynin:1999pq,gusynin2006excitonic,khveshchenko2001magnetic}
\begin{gather}
  \Sigma_{\rm ZLL}(\ii\omega;{\bf r},{\bf r'})
=
-
\int\frac{\dd\Omega}{2\pi}
\,
\mathcal{V}({{\bf r}}-{\bf r}')\,
 G_{\rm ZLL}(\ii\omega+\ii\Omega;{\bf r},{\bf r'}).
\label{eq:Dyson_ZLL}
\end{gather}

 Since the ZLL itself carries no low-energy particle-hole continuum, screening requires virtual transitions to higher Landau levels and is therefore  reduced as \cite{PhysRevLett.87.206401,supplemental}
 \begin{gather}
\Pi_B(0,\vq)=-\frac{N_f}{4\pi\omega_B}q^2 e^{-q^2\ell_B^2/2}. 
 \end{gather}
As a result, when $B$ is small and the IR interaction remains repulsive,  we can seek a self-consistent solution in which interactions dynamically generate a excitonic mass gap $m$ within the ZLL. 
Accordingly, we use the ansatz
$
\widetilde G_{\rm ZLL}(\ii\omega,{\bf q})
=
2\,e^{-q^2\ell_B^2}\,
\frac{\ii\omega\pm m(q)}{\omega^2+m^2(q)}\,
\mathcal P_\pm.
$ Projecting Eq.~\eqref{eq:Dyson_ZLL} onto the ZLL mass channel and assuming $\omega_B \ll \alpha$, we obtain the following result:
\begin{equation}
m(0)
\simeq
\frac{2\omega_B}{N_f}
\ln\!\left(
\frac{2\text{e}^{-\gamma}eB}{\hbar k_p^2}
\right),
\label{eq:m_gap_final}
\end{equation}
where $\gamma \approx 0.5772$ is the Euler constant and $k_p\sim{\omega_p}/{c}$, below which the plasmon mode turns into normal propagating mode.

Thus, under weak magnetic fields, the cavity-induced interaction lifts the degeneracy of the ZLL at any $N_f$. Here, a dynamically generated mass shifts the zero-energy manifold away from the Dirac point $E_{0,\pm}=0$ to $\tilde{E}_{0,\pm}=\mp m(0)$, thereby deforming the originally flat zero-energy level and opening a gap in the ZLL sector. The intervalley scattering can make the split branches equilibrate, and at charge neutrality  the fermions preferentially occupy the lower-energy branch, leading to a charge-density-wave-like sublattice polarization insulating state.

With increasing the magnetic field $B$, screening is progressively weakened, and the effective attractive interaction can be dominant. In this regime, oppositely charged particles and holes can no longer form bound states, but the Landau-level instability occurs in the BCS channel \cite{PhysRevLett.111.046604,PhysRevResearch.3.033260,PhysRevB.96.140506}. For a pair of valleys with the same chirality and Bloch momentum, a nonzero intervalley expectation value $\langle c_{\uparrow,k} c_{\downarrow,-k} \rangle$ is generated, where $\uparrow$ and $\downarrow$ denote the pseudospin freedoms (i.e., two layers of graphene). In the limit $\omega_B\gg \alpha$, the resulting BCS gap is given by~\cite{supplemental}
\begin{equation}
\Delta(0)
\simeq
\frac{\alpha}{2 \pi}
\ln\!\left(
\frac{e^{-\gamma}eB}{\hbar k_p^2}
\right).
\label{eq:Delta_gap_final_main}
\end{equation}

\newsec{Experimental feasibility}
As a possible experimental realization, we discuss the use of graphene nanodisk metasurfaces of Ref.~\cite{fang2014active}, which have the deep-subwavelength patterning relevant to our proposal. The reported disk diameter $D\approx 60~\mathrm{nm}$ and edge-to-edge spacing $g\approx 30~\mathrm{nm}$ imply a lattice period $a\approx 90~\mathrm{nm}$, and the measured plasmonic resonance $\hbar\omega_0\approx 0.17$--$0.18~\mathrm{eV}$ corresponds to $\lambda_0\approx 7~\mu\mathrm{m}$. The resulting ratio $a/\lambda_0\approx 1.3\times10^{-2}\ll1$ places the structure well within the effective uniform medium regime assumed in our analysis. Using $C_s\approx5\times10^{-18}~\mathrm{F}$ with a cavity spacing $d\sim \lambda_0/10\approx 0.7~\mu\mathrm{m}$ and $\varepsilon_d=1$, we have $\hbar\omega_p\approx 0.14~\mathrm{eV}$ and an interaction scale $\alpha\approx 40~\mathrm{meV}$.

These parameters lead to experimentally resolvable many-body energy scales. For example, at $N_f=2$, the above values indicate a zero-field excitonic gap $M(0)\approx 1.2~\mathrm{meV}$, corresponding to a characteristic temperature $T_c\approx 13~\mathrm{K}$.
Under a magnetic field $B=10~\mathrm{T}$ with $v_F=10^6 m/s$, the predicted cavity-induced BCS paring is $\Delta(0)\approx 60~\mathrm{meV}$. 
The key experimental requirement is therefore not extreme geometric or spectral scales, but sufficiently high impedance and weakly dissipative mirrors over the relevant infrared or THz window. Other candidate platforms include split-ring-resonator arrays \cite{yen2004terahertz,scalari2012ultrastrong,scalari2014Superconducting},  superconducting high-impedance surfaces \cite{PhysRevLett.105.247402}, hBN-based phonon-polaritonic structures \cite{caldwell2014sub-diffractional,dai2014tunable}. A complementary route may also be offered by artificial graphene platforms interfaced with microwave resonators \cite{PhysRevLett.111.185307,tarruell2012creating,sterling2014two}, where both cavity coupling and Dirac bands can be controlled in a tunable way.

\newsec{Conclusions}
We have shown that a high-impedance metasurface cavity can qualitatively reshape the low-energy physics of Dirac fermions by engineering the vacuum-mediated electron interaction. The effective interaction drives an excitonic insulating phase for $N_f<N_c$ and induces a non-Fermi-liquid phase for $N_f>N_c$, and lifts the degeneracy of the zeroth Landau level under magnetic fields. From an experimental perspective, these predictions can be tested by the associated depletion or redistribution of low-energy spectral weight; for instance, one may use scanning tunneling spectroscopy to observe suppressed zero-bias conductance \cite{li2009scanning, zhang2009direct} or measure reduced low-frequency spectral weight via infrared or THz conductivity \cite{li2008dirac, basov2011electrodynamics}. Under a magnetic field, one can use Landau-level spectroscopy together with suppressed longitudinal magnetotransport near charge neutrality \cite{novoselov2005two, young2012spin}. More broadly, our work highlights engineered electromagnetic vacuum in metasurface cavities as a new control knob for quantum matter.

\newsec{Acknowledgments}
We thank H.~C. Li and Z. Gong for helpful discussions. 
Y.G. is financially supported by the Global Science Graduate Course (GSGC) program at the University of Tokyo.  
Y.A. acknowledges support from JST FOREST Program (Grant No. JPMJFR222U), JST CREST (Grant No. JPMJCR23I2), and JST [Moonshot R\&D] (Grant No. JPMJMS256J).

\bibliography{ref}


\clearpage
\onecolumngrid
\setcounter{page}{1}

\begin{center}
\textbf{\large Supplemental Material for}\\[1ex]
\textbf{\large ``Cavity-Induced Excitonic Insulation and Non-Fermi-Liquid Behavior in Dirac Materials''}
\end{center}

\vspace{2ex}
\renewcommand{\thefigure}{S\arabic{figure}}
\setcounter{figure}{0}
\renewcommand{\theequation}{S\arabic{equation}}
\setcounter{equation}{0}
\renewcommand{\thesection}{\Roman{section}}
\setcounter{section}{0}
\setcounter{secnumdepth}{4}

\addtocontents{toc}{\protect\setcounter{tocdepth}{0}}
{
\tableofcontents
}
\vspace{2ex}

\vspace{2ex}

\section{Derivation of the cavity-mode dispersion and normalization factor}

In this section, we derive the dispersion relation of the even TM mode in the high-impedance-surface (HIS) cavity and determine the normalization factor $\mathcal N_{\bf q}$ used in the scalar-potential quantization. To this end, we consider two identical high-impedance metasurfaces located at $z=\pm d/2$, separated by a dielectric spacer of relative permittivity $\varepsilon_d$. 
The cavity is translationally invariant in the in-plane coordinate ${\bm r}=(x,y)$.
For a TM mode with in-plane momentum ${\bf q}=q\hat x$, the fields can be written as
\begin{equation}
{\bf E}({\bm r},z,t)=\bigl(E_x(z),0,E_z(z)\bigr)^{\rm T}e^{-\ii{\bf q}\cdot{\bm r}+\ii\omega t},
\qquad
{\bf H}({\bm r},z,t)=\bigl(0,H_y(z),0\bigr)^{\rm T}e^{-\ii{\bf q}\cdot{\bm r}+\ii\omega t}.
\end{equation}
Inside the dielectric spacer, Maxwell's equations give
\begin{equation}
(\partial_z^2-\kappa_d^2)H_y(z)=0,
\qquad
\kappa_d^2=q^2-\varepsilon_d\omega^2/c^2,
\label{eq:Hy_eq_appendix}
\end{equation}
together with
\begin{equation}
E_x(z)=\frac{\ii}{\omega\varepsilon_0\varepsilon_d}\,\partial_z H_y(z),
\qquad
E_z(z)=-\frac{q}{\omega\varepsilon_0\varepsilon_d}\,H_y(z).
\label{eq:ExEz_appendix}
\end{equation}

We focus on the even TM branch, for which $H_y(z)$ is odd under $z\to -z$ and $E_x(z)$ is even. 
A convenient choice is therefore
\begin{equation}
H_y(z)=A_{\bf q}\sinh(\kappa_d z),
\label{eq:Hy_even_appendix}
\end{equation}
which gives
\begin{equation}
E_x(z)=\frac{\ii A_{\bf q}\kappa_d}{\omega\varepsilon_0\varepsilon_d}\cosh(\kappa_d z).
\label{eq:Ex_even_appendix}
\end{equation}
At the upper boundary $z=d/2$, the effective impedance boundary condition reads
\begin{equation}
E_x(d/2)=Z_s(\omega)\,H_y(d/2).
\label{eq:BC_appendix}
\end{equation}
Similarly at $z=-d/2$, we have $E_x(-d/2)=-Z_s(\omega)\,H_y(-d/2),$ since the orientation of the metasurface is different.
Substituting Eqs.~\eqref{eq:Hy_even_appendix} and \eqref{eq:Ex_even_appendix} into Eq.~\eqref{eq:BC_appendix}, we obtain
\begin{equation}
\frac{\ii\kappa_d}{\omega\varepsilon_0\varepsilon_d}
\cosh\!\left(\frac{\kappa_d d}{2}\right)
=
Z_s(\omega)\,
\sinh\!\left(\frac{\kappa_d d}{2}\right),
\end{equation}
or equivalently
\begin{equation}
\coth\!\left(\frac{\kappa_d d}{2}\right)
=
\frac{\omega\varepsilon_0\varepsilon_d\,Z_s(\omega)}{\ii\kappa_d},
\label{eq:dispersion_appendix}
\end{equation}
which is the dispersion relation quoted in the main text.

We now specialize to the deep-subwavelength regime
\begin{equation}
q\gg \omega/c,
\qquad
qd\ll 1,
\end{equation}
for which $\kappa_d\simeq q$ and
\begin{equation}
\coth\!\left(\frac{qd}{2}\right)\simeq \frac{2}{qd}.
\end{equation}
Assuming an LC surface impedance
\begin{equation}
Z_s(\omega)=\left(\ii\omega C_s+\frac{1}{\ii\omega L_s}\right)^{-1},
\end{equation}
Eq.~\eqref{eq:dispersion_appendix} reduces to
\begin{equation}
\frac{2}{qd}
\simeq
\frac{\omega\varepsilon_0\varepsilon_d}{\ii q}
\left(\ii\omega C_s+\frac{1}{\ii\omega L_s}\right)^{-1}.
\end{equation}
After rearranging, this gives
\begin{equation}
\omega_{\bf q}^2\simeq
\omega_p^2
=
\frac{\omega_0^2}{1+\dfrac{\varepsilon_0\varepsilon_d d\,L_s\omega_0^2}{2}},
\qquad
\omega_0=\frac{1}{\sqrt{L_sC_s}},
\label{eq:omegap_appendix}
\end{equation}
showing that the even TM mode becomes nearly dispersionless in the low-energy regime.

Because the mode is quasielectrostatic, it is convenient to introduce a scalar potential $\phi$ such that ${\bf E}\simeq -\nabla\phi$. 
For the even TM mode, the potential profile is proportional to $\cosh(qz)$, and we can write
\begin{equation}
\hat\phi({\bm r},z)
=
\sum_{\bf q}
\frac{1}{\sqrt{\mathcal N_{\bf q}}}
\cosh(qz)\,e^{-\ii{\bf q}\cdot{\bm r}}
\left(\hat a_{\bf q}+\hat a^\dagger_{-{\bf q}}\right).
\label{eq:phi_ansatz_appendix}
\end{equation}
The corresponding electric field is
\begin{equation}
\hat E_x=-\partial_x\hat\phi,
\qquad
\hat E_z=-\partial_z\hat\phi.
\end{equation}
In the narrow-gap limit $qd\ll1$, one has
\begin{equation}
\frac{|E_z|}{|E_x|}
\sim q|z|
\ll1,
\end{equation}
so the mode is dominated by its in-plane electric field on the plane near the cavity center.

To fix the normalization, we evaluate the total energy stored in one mode. A complete 
calculation should involve summing the time-averaged energies of both the 
capacitance and the inductance; however, for the convenience of our 
calculation, the normalization is determined by calculating the energy when 
the capacitance reaches its maximum energy storage, which should be the same as the total energy in electric field and magnetic field.
There are two contributions: the electric-field energy in the dielectric slab and the reactive capacitive energy stored on the two HIS boundaries. 
To leading order in $qd$, the field energy is
\begin{equation}
U_{\rm diel}
=
\frac{\varepsilon_0\varepsilon_d}{2}
\int_{-d/2}^{d/2}\dd z
\int_S \dd^2r\,
|\nabla\phi|^2
\simeq
\frac{S q^2}{\mathcal N_{\bf q}}
\left(\frac{\varepsilon_0\varepsilon_d d}{2}\right)
\left(\hat a_{\bf q}^\dagger\hat a_{\bf q}+\frac12\right),
\label{eq:Udiel_appendix}
\end{equation}
where $S$ is the in-plane quantization area and terms like $\hat{a}_q\hat{a}_{-q}$ are canceled by the magnetic part in the inductance.
The reactive energy stored on the two identical high-impedance boundaries is
\begin{equation}
U_{\rm surf}
=
\frac{S q^2}{\mathcal N_{\bf q}}\,C_s
\left(\hat a_{\bf q}^\dagger\hat a_{\bf q}+\frac12\right).
\label{eq:Usurf_appendix}
\end{equation}
Adding the two contributions gives
\begin{equation}
U_{\bf q}
=
\frac{S q^2}{\mathcal N_{\bf q}}
\left(C_s+\frac{\varepsilon_0\varepsilon_d d}{2}\right)
\left(\hat a_{\bf q}^\dagger\hat a_{\bf q}+\frac12\right).
\label{eq:Uq_appendix}
\end{equation}
Imposing canonical quantization,
\begin{equation}
U_{\bf q}
=
\hbar\omega_{\bf q}
\left(\hat a_{\bf q}^\dagger\hat a_{\bf q}+\frac12\right),
\end{equation}
we obtain
\begin{equation}
\mathcal N_{\bf q}
=
\frac{S q^2}{\hbar\omega_{p}}
\left(C_s+\frac{\varepsilon_0\varepsilon_d d}{2}\right).
\label{eq:Nq_appendix}
\end{equation}
Substituting Eq.~\eqref{eq:Nq_appendix} into Eq.~\eqref{eq:phi_ansatz_appendix}, one finally arrives at
\begin{equation}
\hat{\phi}({\bm r},z)
=
\sum_{\bf q}
\sqrt{\frac{\hbar\omega_{p}}{S q^2\left(C_s+\dfrac{\varepsilon_0\varepsilon_d d}{2}\right)}}
\,\cosh(qz)\,e^{-\ii{\bf q}\cdot{\bm r}}
\left(\hat a_{\bf q}+\hat a^\dagger_{-{\bf q}}\right),
\end{equation}
which is the form used in the main text.

\section{Derivation of the excitonic gap equation and mass scale}
\label{app:gap_derivation}
\subsection{Frequency integration with the retarded kernel and the cutoff structure}

We now analyze how the full retarded cavity kernel determines the natural energy and momentum window of the excitonic gap equation before the static reduction used in the main text.
The key point is that the finite cavity mode frequency $\omega_p$ does not merely set an overall interaction scale, but also restricts the internal fermionic energies that can participate efficiently in pairing.

The dressed Green's function satisfies
\begin{equation}
G^{-1}(\ii\omega,{\bf k})
=
G_0^{-1}(\ii\omega,{\bf k})
-
\Sigma(\ii\omega,{\bf k}),
\label{eq:app_Dyson}
\end{equation}
with the free propagator
\begin{gather}
G_0(\ii\omega,{\bf k})
=
\frac{\ii\omega+v_F\bm\sigma\!\cdot\!{\bf k}}
{\omega^2+v_F^2k^2}.
\label{eq:app_G0}
\end{gather}
In the symmetry-broken phase, we project the self-energy onto the mass channel and approximate
\begin{equation}
\Sigma(\ii\omega,{\bf k})\simeq M(k)\sigma_z ,
\label{eq:app_sigma_mass}
\end{equation}
where $\sigma_z$ acts on the sublattice space and corresponds to $\gamma_0$ in the covariant notation. 
The Dyson-Schwinger equation then takes the form
\begin{equation}
\Sigma(\ii\omega,{\bf k})
=
-e^2\int\frac{\dd\Omega}{2\pi}
\int\frac{\dd^2q}{(2\pi)^2}\,
 D_{\phi}(\ii\Omega,{\bf q})\,
G(\ii\omega+\ii\Omega,{\bf k}+{\bf q}).
\label{eq:app_rainbow}
\end{equation}

Keeping the full frequency dependence of the cavity-mediated interaction, the Dyson-Schwinger equation in the mass channel takes the form
\begin{equation}
M(p)
=
\int \frac{\dd^2k}{(2\pi)^2}
\int \frac{\dd\Omega}{2\pi}
\,
\frac{2\omega_p |g_q|^2}
{\Omega^2+{\omega}_p^2}
\,
\frac{M(k)}
{\Omega^2+v_F^2k^2+M^2(k)},
\label{eq:SDE_retarded_full}
\end{equation}
where
\begin{equation}
q\equiv |\vp-\vk|,
\qquad
E_k\equiv \sqrt{v_F^2k^2+M^2(k)}.
\end{equation}
The gap equation reduces to
\begin{equation}
M(p)
=
\int \frac{\dd^2k}{(2\pi)^2}\,M(k)\,I(\vp,\vk),
\label{eq:M_with_I_appendix}
\end{equation}
with
\begin{equation}
I(\vp,\vk)
=
\int \frac{\dd\Omega}{2\pi}
\,
\frac{2\omega_p |g_q|^2}
{\Omega^2+{\omega}_p^2}
\,
\frac{1}{\Omega^2+E_k^2}.
\label{eq:I_def_appendix}
\end{equation}

Using the standard identity
\begin{equation}
\int \frac{\dd\Omega}{2\pi}
\frac{1}{(\Omega^2+a^2)(\Omega^2+b^2)}
=
\frac{1}{2ab(a+b)},
\qquad a,b>0,
\label{eq:integral_identity_appendix}
\end{equation}
we obtain
\begin{equation}
I(\vp,\vk)
=
|g_q|^2
\frac{1}{E_k}
\frac{1}{(E_k+{\omega}_p)}.
\label{eq:I_result_appendix}
\end{equation}
Therefore the retarded gap equation becomes
\begin{equation}
M(p)
=
\int \frac{\dd^2k}{(2\pi)^2}
\,
|g_q|^2
\frac{M(k)}
{E_k\left(E_k+{\omega}_p\right)}.
\label{eq:gap_retarded_after_freq_appendix}
\end{equation}

Equation~\eqref{eq:gap_retarded_after_freq_appendix} makes two facts explicit:
first, dynamical screening enters through ${\omega}_p$; second, retardation produces the additional suppression factor $(E_k+{\omega}_p)^{-1}$.

In the low-energy regime relevant to the gap-opening instability, the dominant contribution comes from fermionic states with
\begin{equation}
E_k\ll  \omega_p.
\label{eq:adiabatic_condition_appendix}
\end{equation}
In this regime one may approximate
\begin{equation}
\frac{1}{E_k+{\omega}_p}\simeq \frac{1}{{\omega}_p},
\label{eq:low_energy_reduction_appendix}
\end{equation}
where the second relation uses the weak dispersion of the cavity mode in the quasielectrostatic regime.
Equation~\eqref{eq:gap_retarded_after_freq_appendix} then reduces to
\begin{equation}
M(p)
\simeq
\int \frac{\dd^2k}{(2\pi)^2}
\,
\frac{|g_q|^2}{{\omega}_q}
\frac{M(k)}{E_k}.
\label{eq:gap_static_intermediate_appendix}
\end{equation}
Using the static interaction in the form
\begin{equation}
\mathcal{V}(q,0)=\frac{2|g_q|^2}{{\omega}_p},
\label{eq:V_static_screened_appendix}
\end{equation}
we obtain the static gap equation
\begin{equation}
M(p)
=
\int \frac{\dd^2k}{(2\pi)^2}\,
\mathcal{V}(q,0)\,
\frac{M(k)}{2E_k},
\label{eq:gap_static_final_appendix}
\end{equation}
which is the equation used in the main text.

Equation~\eqref{eq:gap_retarded_after_freq_appendix} also makes clear why the finite cavity resonance provides a natural ultraviolet energy cutoff.
For $E_k\gg {\omega}_p$, the retarded factor behaves as
\begin{equation}
\frac{1}{E_k(E_k+{\omega}_q)}
\simeq
\frac{1}{E_k^2},
\qquad
E_k\gg {\omega}_p,
\label{eq:high_energy_suppression_appendix}
\end{equation}
so the kernel is parametrically suppressed compared with the low-energy form $1/(E_k\widetilde{\omega}_q)$.
Physically, once the internal fermionic energy exceeds the cavity resonance, the interaction is strongly reduced by retardation: the bosonic mode can no longer mediate pairing efficiently at such high frequencies.
Therefore the relevant internal energy window is effectively restricted to
\begin{equation}
E_k\lesssim \omega_p,
\label{eq:energy_cutoff_appendix}
\end{equation}
which identifies $\omega_p$ as the natural ultraviolet cutoff of the low-energy gap equation.

The same reasoning determines the momentum window relevant to the mass self-energy $\Sigma_0$.
In the pairing kernel, the fermionic energy entering Eq.~\eqref{eq:gap_retarded_after_freq_appendix} is $E_k\simeq v_Fk$ as long as $v_Fk\gg M(k)$.
The retardation-induced suppression therefore becomes important once
\begin{equation}
v_Fk\sim \omega_p.
\label{eq:vFk_omegap_appendix}
\end{equation}
Accordingly, the ultraviolet momentum scale of the low-energy gap equation is
\begin{equation}
k_{\rm UV}\sim \frac{\omega_p}{v_F}.
\label{eq:kUV_appendix}
\end{equation}
Beyond this scale, the fermionic states are too energetic to couple efficiently through the nearly flat cavity mode, and the static reduction leading to Eq.~\eqref{eq:gap_static_final_appendix} is no longer justified.

There is also a lower momentum boundary associated with the validity of the quasielectrostatic cavity mode itself.
The cavity interaction used throughout the main text relies on the deep-subwavelength condition
\begin{equation}
q\gg \omega/c.
\label{eq:quasistatic_condition_appendix}
\end{equation}
Since the dominant retarded pairing processes are restricted to frequencies of order $\omega\lesssim \omega_p$, the lower edge of the momentum window is set parametrically by
\begin{equation}
q_{\rm IR}\sim \frac{\omega_p}{c}.
\label{eq:qIR_appendix}
\end{equation}
Equivalently, for the momentum integration in the gap equation one may regard
\begin{equation}
k_{\rm IR}\sim \frac{\omega_p}{c}
\label{eq:kIR_appendix}
\end{equation}
as the infrared cutoff associated with the breakdown of the quasistatic plasmonic description at too small momentum transfer.
This cutoff is not generated by the fermionic dynamics itself, but by the range of validity of the effective cavity kernel.

Taken together, Eqs.~\eqref{eq:kUV_appendix} and \eqref{eq:kIR_appendix} show that the retarded HIS-mediated interaction is effective only within the momentum window
\begin{equation}
\frac{\omega_p}{c}
\;\lesssim\;
k
\;\lesssim\;
\frac{\omega_p}{v_F}.
\label{eq:k_window_appendix}
\end{equation}
Within this window the interaction may be treated as quasistatic, whereas outside it either retardation suppresses pairing at large $k$ or the quasielectrostatic cavity-mode approximation breaks down at small $k$.


\subsection{Vacuum Polarization of 2+1D Dirac Fermions at Zero Magnetic Field}

In the absence of a magnetic field, the one-loop vacuum polarization tensor for $N_f$ flavors of Dirac fermions is given by the standard Feynman rule:
\begin{equation}
\Pi^{\mu\nu}(p) = -\ii N_f \int \frac{\dd^3k}{(2\pi)^3} \Tr \left[ \gamma^\mu S(k+p) \gamma^\nu S(k) \right],
\end{equation}
where the free fermion propagator is $S(k) = (\slashed{k} + m)/(k^2 - m^2 + \ii\epsilon)$. 

Evaluating the trace of the $2 \times 2$ Dirac matrices in 2+1 dimensions, $\Tr(\gamma^\mu\gamma^\nu) = 2\eta^{\mu\nu}$ and $\Tr(\gamma^\mu\gamma^\alpha\gamma^\nu\gamma^\beta) = 2(\eta^{\mu\alpha}\eta^{\nu\beta} - \eta^{\mu\nu}\eta^{\alpha\beta} + \eta^{\mu\beta}\eta^{\nu\alpha})$, we obtain:
\begin{equation}
\Tr[\gamma^\mu(\slashed{k}+m)\gamma^\nu(\slashed{k}+m)] = 2 \left[ (k+p)^\mu k^\nu + (k+p)^\nu k^\mu - \eta^{\mu\nu} (k \cdot (k+p) - m^2) \right].
\end{equation}

To perform the momentum integration, we introduce the Feynman parameter $x \in [0,1]$ and shift the loop momentum $l = k + xp$. Dropping terms linear in $l$ due to symmetric integration, the tensor takes the form:
\begin{equation}
\Pi^{\mu\nu}(p) = -2\ii N_f \int_0^1 \dd x \int \frac{\dd^3l}{(2\pi)^3} \frac{2l^\mu l^\nu - \eta^{\mu\nu} l^2 - 2x(1-x)p^\mu p^\nu + \eta^{\mu\nu}(x(1-x)p^2 + m^2)}{[l^2 - \Delta + \ii\epsilon]^2},
\end{equation}
where $\Delta = m^2 - x(1-x)q^2$. Using dimensional regularization to evaluate the integrals, the result naturally satisfies the Ward identity $q_\mu \Pi^{\mu\nu} = 0$ and can be decomposed as:
\begin{equation}
\Pi^{\mu\nu}(q) = (q^2 \eta^{\mu\nu} - q^\mu q^\nu) \Pi(q^2).
\end{equation}

The scalar polarization function $\Pi(p)$ is found to be:
\begin{equation}
\Pi(p) = \frac{N_f}{2\pi} \int_0^1 \dd x \frac{x(1-x)}{\sqrt{m^2 - x(1-x)p^2}}.
\end{equation}
For massless Dirac fermions ($m=0$), which is relevant to, e.g., graphene, the static polarization function simplifies to
\begin{equation}
\Pi(p) = \frac{N_f}{16 P},\;\;\;\;P = \sqrt{-p^2}.
\end{equation}
This $1/P$ dependence is a hallmark of 2+1-dimensional massless Dirac fermions, leading to a strongly momentum-dependent dielectric screening.
In condensed matter physics, particularly when studying Coulomb screening and dielectric properties, the most relevant quantity is the \emph{density-density response function}. In the field-theoretic language, the density operator is $\rho = \bar{\psi} \gamma^0 \psi$, which means the density-density correlation is exactly given by the time-time component ($\mu = \nu = 0$) of the polarization tensor:
\begin{equation}
\Pi^{00}(p) = (p^2 \eta^{00} -p^0 p^0) \Pi(p).
\end{equation}
Using the Minkowski metric $\eta^{00} = 1$ and the four-momentum $p^\mu = (\omega, \mathbf{q})$, we have $p^2 = \omega^2 - {q}^2$. Substituting this into the 00-component yields:
\begin{equation}
\Pi^{00}(\omega, \mathbf{q}) = (\omega^2 - {q}^2 - \omega^2) \Pi(q^2) = -{q}^2 \frac{N_f}{16 \sqrt{{q}^2 - \omega^2}}.
\end{equation}

To evaluate the thermodynamic properties and Matsubara frequency responses, we perform a Wick rotation to Euclidean space by taking $\omega \to \ii\omega$. The momentum squared becomes strictly negative. In the Euclidean signature, the retarded density-density response function $\Pi(\mathbf{q}, \ii\omega)$ is defined with an overall Wick rotation to the Minkowski $\Pi^{00}$, leading to:
\begin{equation}
\Pi( \ii\omega, \mathbf{q}) = -\frac{N_f {q}^2}{16 \sqrt{\omega^2 + v_F^2{q}^2}}.
\label{eq:Pi_vF1}
\end{equation}

We next derive the low-energy gap equation used in the main text and the resulting scaling form of the excitonic mass. 
We work within the rainbow Dyson approximation and use the static RPA-screened interaction appropriate to the low-energy instability. 
The basic logic is as follows: starting from the self-consistent Dyson-Schwinger equation, we first reduce the problem to a nonlinear integral equation for the momentum-dependent mass function. We then perform the angular integration and convert the result into a local differential equation, and finally solve the latter subject to infrared and ultraviolet matching conditions.

We thus replace the full propagator by the RPA-screened interaction 
\begin{equation}
\tilde {\mathcal{V}}_{\rm RPA}(\ii\omega,{\bf q})
=
\frac{{\mathcal{V}}_0(\ii\omega,{\bf q})}
{1-\Pi(\ii\omega,{\bf q}){\mathcal{V}}_0(\ii\omega,{\bf q})}.
\label{eq:app_RPA}
\end{equation}
For the low-energy cavity mode, the bare static interaction has the infrared form
\begin{equation}
{\mathcal{V}}_0({\bf q})\simeq -\frac{\alpha}{q^2},
\label{eq:app_V0}
\end{equation}
while the static polarization of massless Dirac fermions is
\begin{equation}
\Pi({\bf q},0)=-\frac{N_f q}{16v_F}.
\label{eq:app_Pi0}
\end{equation}
Substituting Eqs.~\eqref{eq:app_V0} and \eqref{eq:app_Pi0} into Eq.~\eqref{eq:app_RPA} gives
\begin{equation}
{\mathcal{V}}_{\rm RPA}(0,{\bf q})
=
\frac{-\alpha/q^2}{1-\dfrac{N_f q}{16v_F}\dfrac{\alpha}{q^2}}
\label{eq:app_VRPA_static}
\end{equation}
In the infrared regime $q\ll \lambda_{\rm scr}$, this reduces to
\begin{equation}
{\mathcal{V}}_{\rm RPA}(0,{\bf q})\simeq \frac{16v_F}{N_f}\frac{1}{q}.
\label{eq:app_Vinfrared}
\end{equation}
Thus screening softens the bare logarithmic interaction but preserves a singular long-range kernel and the UV cutoff is set arround $\Lambda_{\rm UV}=\alpha$, which is the origin of the nonperturbative instability. 

In practice, the ultraviolet cutoff $k_{\text{UV}}$ should be determined by $\min[\omega_p/v_F, 2\pi/d]$. While $1/d$ represents a geometric constraint contingent upon the specific cavity architecture—which could be further increased via optimized fabrication—the scale $\alpha/v_F$ constitutes the physical energy scale where the effective field theory remains valid. Consequently, we adopt $k_{UV}=\alpha/v_F$ as the UV cutoff for our theory.

\subsection{Detailed Derivation of the Gap Equation }

In this section, we provide the full derivation of the self-consistent gap equation and the resulting scaling behavior. We start by projecting the rainbow-approximation Dyson-Schwinger equation onto the mass channel. And in the low frequency window we care about we can approximate the photon by static potential,

\begin{equation}
M(p)
=
\int\frac{\dd^2 p}{(2\pi)^2}\int\frac{\dd \Omega}{2\pi}\;
\mathcal{V}(|\vk-\vp|)\;
\frac{M(k)}{\Omega^2+v_F^2k^2+M^2(k)}.
\label{eq:DMG_SD_start}
\end{equation}

After performing the internal frequency integration, we obtain the zero-frequency nonlinear integral equation for the mass function $M({\bf p})$:

\begin{equation}
M({ p}) = \int\frac{\dd^2k}{(2\pi)^2}\,  {\mathcal{V}}_{\rm RPA}(0,|{\bf p}-{\bf k}|)\, \frac{M(k)}{2E_k}, \qquad E_k=\sqrt{v_F^2k^2+M^2(k)}.
\label{eq:app_gap_integral0}
\end{equation}
Where we use the infrared limit of the screened potential ${\mathcal{V}}_{\rm RPA}$ from Eq.~\eqref{eq:app_Vinfrared}, the kernel is dominated by the cavity-induced $1/q$ term, leading to
\begin{equation}
M({\bf p}) \simeq \frac{16v_F}{N_f} \int\frac{\dd^2k}{(2\pi)^2}\, \frac{1}{|{\bf p}-{\bf k}|}\, \frac{M(k)}{2E_k}.
\label{eq:app_gap_integral1}
\end{equation}

After performing the internal frequency integration, we obtain the zero-frequency nonlinear integral equation for the mass function $M({\bf p})$:
\begin{equation}
M({\bf p}) = \int\frac{\dd^2k}{(2\pi)^2}\,  {\mathcal{V}}_{\rm RPA}(0,|{\bf p}-{\bf k}|)\, \frac{M(k)}{2E_k}, \qquad E_k=\sqrt{v_F^2k^2+M^2(k)}.
\label{eq:app_gap_integral0}
\end{equation}
Using the infrared limit of the screened potential ${\mathcal{V}}_{\rm RPA}$ from Eq.~\eqref{eq:app_Vinfrared}, the kernel is dominated by the cavity-induced $1/q$ term, leading to
\begin{equation}
M({\bf p}) \simeq \frac{16v_F}{N_f} \int\frac{\dd^2k}{(2\pi)^2}\, \frac{1}{|{\bf p}-{\bf k}|}\, \frac{M(k)}{2E_k}.
\label{eq:app_gap_integral1}
\end{equation}
To evaluate the integration, we express the measure as $\dd^2k=k\,\dd k\,\dd\theta$ and expand the distance as $|{\bf p}-{\bf k}|=\sqrt{p^2+k^2-2pk\cos\theta}$, which gives
\begin{equation}
M(p) = \frac{8v_F}{N_f} \int_0^\alpha \frac{k\,\dd k}{(2\pi)^2}\, \frac{M(k)}{E_k} \int_0^{2\pi}\dd\theta\, \frac{1}{\sqrt{p^2+k^2-2pk\cos\theta}}.
\label{eq:app_gap_integral2}
\end{equation}
To proceed analytically, we adopt the standard angle-averaging approximation, which captures both the  UV and IR scalings correctly \cite{Appelquist:1988sr}:
\begin{equation}
\frac{1}{2\pi}\int_0^{2\pi}\dd\theta\, \frac{1}{|{\bf p}-{\bf k}|} \;\approx\; \frac{1}{\max(p,k)}.
\label{eq:app_angle_avg}
\end{equation}
Substituting this into Eq.~\eqref{eq:app_gap_integral2}, we reduce the two-dimensional integral to a one-dimensional form:
\begin{equation}
M(p) = \frac{4v_F}{\pi N_f} \left[ \frac{1}{p}\int_0^p \dd k\,\frac{k\,M(k)}{E_k} + \int_p^\alpha \dd k\,\frac{M(k)}{E_k} \right].
\label{eq:app_gap_1d}
\end{equation}
By defining the auxiliary functions $F(k)\equiv \frac{k\,M(k)}{E_k}$ and $G(k)\equiv \frac{M(k)}{E_k}$, the integral equation can be written compactly as
\begin{equation}
M(p) = \frac{4v_F}{\pi N_f} \left[ \frac{1}{p}\int_0^p \dd k\,F(k) + \int_p^\alpha \dd k\,G(k) \right].
\label{eq:app_gap_FG}
\end{equation}

Next, we convert this integral equation into a local differential equation. Multiplying Eq.~\eqref{eq:app_gap_FG} by $p$ yields
\begin{equation}
p\,M(p) = \frac{4v_F}{\pi N_f} \left[ \int_0^p \dd k\,F(k) + p\int_p^\alpha \dd k\,G(k) \right].
\label{eq:app_pM}
\end{equation}
Differentiating once with respect to $p$ using the Leibniz rule, we find
\begin{equation}
\frac{\dd}{\dd p}[pM(p)] = \frac{4v_F}{\pi N_f} \left[ F(p)+\int_p^\alpha \dd k\,G(k)-pG(p) \right].
\label{eq:app_first_derivative}
\end{equation}
Given that $F(p)=pG(p)$ by definition, the boundary terms cancel, simplifying the expression to
\begin{equation}
\frac{\dd}{\dd p}[pM(p)] = \frac{4v_F}{\pi N_f} \int_p^\alpha \dd k\,G(k).
\label{eq:app_first_derivative2}
\end{equation}
Applying a second derivative and utilizing the identity $\frac{\dd^2}{\dd p^2}[pM(p)] = \frac{1}{p}\frac{\dd}{\dd p} \left( p^2\frac{\dd M(p)}{\dd p} \right)$, we obtain the exact nonlinear local equation:
\begin{equation}
\frac{\dd}{\dd p} \left( p^2\frac{\dd M(p)}{\dd p} \right) = -\frac{4v_F}{\pi N_f}\, \frac{p\,M(p)}{\sqrt{v_F^2p^2+M^2(p)}}.
\label{eq:app_nonlinear_DE}
\end{equation}
In the scaling regime $M(0)\ll p\ll \alpha$, where $M(p)\ll v_F p$, the denominator can be linearized to $v_F p$. Equation~\eqref{eq:app_nonlinear_DE} then reduces to the Euler-type differential equation
\begin{equation}
\frac{\dd}{\dd p} \left( p^2\frac{\dd M(p)}{\dd p} \right) + \frac{4}{\pi N_f}\,M(p) = 0.
\label{eq:app_linear_DE}
\end{equation}
Assuming a power-law solution $M(p)\sim p^s$, the characteristic equation and its roots are
\begin{gather}
s(s+1)+\frac{4}{\pi N_f}=0, \\
s=-\frac{1}{2}\pm\sqrt{\frac{1}{4}-\frac{4}{\pi N_f}}.
\end{gather}
From this, we identify the critical flavor number $N_c=16/\pi$ as the threshold where the roots become complex.

The existence of physically viable solutions is determined by the boundary conditions. The  IR condition follows from the regularity of the mass function at the origin, $p^2\frac{\dd M(p)}{\dd p}\Big|_{p=M(0)}=0$. The  UV condition is obtained by evaluating Eq.~\eqref{eq:app_first_derivative2} at the cutoff $p=\alpha$:
\begin{equation}
\left[ M(p)+p\frac{\dd M(p)}{\dd p} \right]_{p=\alpha}=0.
\label{eq:app_UV_bc}
\end{equation}
For $N_f<N_c$, we define $\gamma=\frac12\sqrt{\frac{N_c}{N_f}-1}$ and express the general oscillatory solution as
\begin{equation}
M(p) = A\,p^{-1/2} \sin\!\left[ \gamma\ln\!\left(\frac{p}{M(0)}\right)+\delta_{N_f} \right].
\label{eq:app_M_solution}
\end{equation}
Differentiating this solution to satisfy the IR condition, we find
\begin{equation}
p^2\frac{\dd M}{\dd p} = A\,p^{1/2} \left[ -\frac12 \sin\!\left( \gamma\ln\frac{p}{M(0)}+\delta_{N_f} \right) + \gamma \cos\!\left( \gamma\ln\frac{p}{M(0)}+\delta_{N_f} \right) \right].
\end{equation}
At $p=M(0)$, this yields the phase shift $\tan\delta_{N_f}=2\gamma$. Finally, substituting the solution into the UV boundary condition \eqref{eq:app_UV_bc} gives
\begin{equation}
\tan\!\left( \gamma\ln\frac{\alpha}{M(0)}+\delta_{N_f} \right) = -2\gamma.
\label{eq:app_tan_condition}
\end{equation}
For the ground state, we match the argument to the first node, $\gamma\ln\frac{\alpha}{M(0)}+\delta_{N_f} \simeq \pi$. This leads to the final Miransky scaling for the mass gap:
\begin{equation}
M(0) \simeq \alpha \exp\!\left[ -\frac{\pi-\delta_{N_f}}{\gamma} \right] \sim \alpha \exp\!\left[ -\frac{2\pi}{\sqrt{N_c/N_f-1}} \right].
\end{equation}
This exponential behavior confirms that the excitonic insulation occurs via an infinite-order quantum phase transition at $N_c$.

\section{Derivation of the Non-Fermi-Liquid Behavior}

For fermion flavors $N_f > N_c$, the gap equation does not admit a non-trivial solution, meaning the ground state remains gapless ($M=0$). However, the strong, long-range cavity-mediated interaction fundamentally reconstructs the low-energy excitations. In this section, we derive the non-Fermi-liquid behavior by evaluating the one-loop self-energy. We will demonstrate that the highly singular nature of the effective interaction leads to an anomalous power-law scaling of the Fermi velocity and a complete suppression of the quasiparticle residue, signaling the breakdown of the standard Fermi liquid paradigm.

\subsection{One-loop self-energy}

In the gapless phase, the low-energy screened interaction is polarization dominated and takes the retarded form
\begin{equation}
\tilde {\mathcal{V}}_{\rm RPA}(\ii\Omega,{\bf q})
\simeq
\frac{16 }{N_f \tilde{Z}^2 }\,
\frac{\sqrt{\Omega^2+\tilde{v}_F^2q^2}}{q^2},
\label{eq:app_V_gapless}
\end{equation}
valid within the low-energy window $|\Omega|\lesssim \alpha$. 
The one-loop self-energy is
\begin{equation}
\Sigma(\ii\omega,{\bf k})
=
-\int\frac{\dd\Omega}{2\pi}
\int\frac{\dd^2q}{(2\pi)^2}\,
\tilde {\mathcal{V}}_{\rm RPA}(\ii\Omega,{\bf q})\,
G(\ii\omega+\ii\Omega,{\bf k}+{\bf q}).
\label{eq:app_Sigma_start}
\end{equation}
Substituting Eq.~\eqref{eq:app_Dyson} gives
\begin{equation}
\Sigma(\ii\omega,{\bf k})
=
-\int\frac{\dd\Omega}{2\pi}
\int\frac{\dd^2q}{(2\pi)^2}\,
\tilde {\mathcal{V}}_{\rm RPA}(\ii\Omega,{\bf q})\,
\frac{\ii(\omega+\Omega)+\tilde{v}_F\,\boldsymbol{\sigma}\!\cdot\!({\bf k}+{\bf q})}
{(\omega+\Omega)^2+\tilde{v}_F^2|{\bf k}+{\bf q}|^2}.
\label{eq:app_Sigma_expanded}
\end{equation}
We parameterize the self-energy as
\begin{equation}
\Sigma(\ii\omega,{\bf k})
=
\tilde{Z}^{-1}[\ii\omega\,\Sigma_\omega(\omega,k)
+
\tilde{v}_F\,\boldsymbol{\sigma}\!\cdot\!{\bf k}\,\Sigma_k(\omega,k)]
\label{eq:app_Sigma_decomp}
\end{equation}
With this convention, the dressed inverse Green's function takes the form
\begin{equation}
G^{-1}(\ii\omega,{\bf k})
=
\tilde{Z}^{-1}\left[-\ii\omega(1+\Sigma_\omega(\omega,k))
+
\tilde{v}_F(1-\Sigma_k(\omega,k)\bigr)
\boldsymbol{\sigma}\!\cdot\!{\bf k}\right].
\label{eq:app_Ginv_dressed}
\end{equation}

\subsection{Derivation of $\Sigma_\omega$}

The temporal coefficient is extracted by differentiation with respect to the external frequency,
\begin{equation}
\Sigma_\omega(\mu)
=
\frac{1}{2}
\Tr\left[
\frac{\partial\Sigma(\ii\omega,{\bf 0})}{\partial(\ii\omega)}
\right]_{\omega=0},
\label{eq:app_proj_omega}
\end{equation}
where the trace is over the two-component pseudospin space. 
Differentiating Eq.~\eqref{eq:app_Sigma_expanded} and setting ${\bf k}=0$ gives
\begin{equation}
\Sigma_\omega
=
-\int\frac{\dd\Omega}{2\pi}
\int\frac{\dd^2q}{(2\pi)^2}\,
\tilde {\mathcal{V}}_{\rm RPA}(\ii\Omega,{\bf q})
\frac{\tilde{v}_F^2q^2-\Omega^2}{(\Omega^2+\tilde{v}_F^2q^2)^2}.
\label{eq:app_Sigma_omega_integral}
\end{equation}
Substituting Eq.~\eqref{eq:app_V_gapless}, we obtain
\begin{equation}
\Sigma_\omega
=
-\frac{16}{N_f }
\int\frac{\dd\Omega}{2\pi}
\int\frac{\dd^2q}{(2\pi)^2}\,
\frac{\tilde{v}_F^2q^2-\Omega^2}
{q^2(\Omega^2+\tilde{v}_F^2q^2)^{3/2}}.
\label{eq:app_Sigma_omega_V}
\end{equation}
Introducing $x=\Omega/(\tilde{v}_F q)$, so that $\dd\Omega\dd q=\tilde{v}_F q\,\dd x\dd q$, yields
\begin{equation}
\Sigma_\omega(\mu)
=
-\frac{8}{\pi^2N_f}
\int^{\alpha/\tilde{v}_F}_{\mu/\tilde{v}_F}\frac{\dd q}{q}
\int_0^{\alpha/(\tilde{v}_Fq)}\dd x\,
\frac{1-x^2}{(1+x^2)^{3/2}}.
\label{eq:app_Sigma_omega_x}
\end{equation}
For large upper cutoff $X=\alpha/(\tilde{v}_Fq)\gg1$,
\begin{equation}
\int_0^{X}\dd x\,
\frac{1-x^2}{(1+x^2)^{3/2}}
=
2-\operatorname{arcsinh}X+\mathcal O(X^{-1})
\simeq
-\ln X+2-\ln2.
\label{eq:app_xint_omega}
\end{equation}
The remaining momentum integral is therefore of the form $\int (\dd q/q)\ln(\alpha/\tilde{v}_F q)$ and generates a double logarithm,
\begin{equation}
\Sigma_\omega(\mu)
\simeq
\frac{4}{\pi^2N_f}\,
\ln^2\!\frac{\alpha}{\mu}-\frac{8(2-\ln 2)}{\pi^2N_f}\ln{\frac{\alpha}{\mu}}.
\label{eq:app_Sigma_omega_result}
\end{equation}

In the  IR limit, where the running energy scale $\mu \ll \alpha$, the asymptotic expansion of the integral correctly captures the dependence of $\mu$ and  $\Sigma(\mu)$.

\subsection{Derivation of $\Sigma_k$}

Starting from
\begin{equation}
\Sigma(\ii\omega,{\bf k})
=
-\int\frac{\dd\Omega}{2\pi}
\int\frac{\dd^2q}{(2\pi)^2}\,
\tilde {\mathcal{V}}_{\rm RPA}(\ii\Omega,{\bf q})\,
\frac{\ii(\omega+\Omega)+\tilde{v}_F\,\boldsymbol{\sigma}\!\cdot\!({\bf k}+{\bf q})}
{(\omega+\Omega)^2+\tilde{v}_F^2|{\bf k}+{\bf q}|^2},
\label{eq:Sigma_start_spatial_appendix}
\end{equation}
we extract the spatial coefficient using
\begin{equation}
\Sigma_k(\mu)
=
\sum_{i=x,y}\frac{1}{4\tilde{v}_F}
\Tr\left[
\sigma_i
\frac{\partial\Sigma(0,{\bf k})}{\partial k_i}
\right]_{|{\bf k}|=0}.
\label{eq:proj_k_appendix}
\end{equation}
At $\omega=0$, differentiation with respect to $k_i$ gives
\begin{equation}
\frac{\partial}{\partial k_i}
\frac{\ii\Omega+\tilde{v}_F\,\boldsymbol{\sigma}\!\cdot\!({\bf k}+{\bf q})}
{\Omega^2+\tilde{v}_F^2|{\bf k}+{\bf q}|^2}
=
\frac{\tilde{v}_F\sigma_i}{D}
-
\frac{2\tilde{v}_F^2(k_i+q_i)\bigl(\ii\Omega+\tilde{v}_F\,\boldsymbol{\sigma}\!\cdot\!({\bf k}+{\bf q})\bigr)}{D^2},
\qquad
D=\Omega^2+\tilde{v}_F^2|{\bf k}+{\bf q}|^2.
\label{eq:dSigma_dk_appendix}
\end{equation}
Setting ${\bf k}=0$, we obtain
\begin{equation}
\left.
\frac{\partial\Sigma(0,{\bf k})}{\partial k_i}
\right|_{{\bf k}=0}
=
-\int\frac{\dd\Omega}{2\pi}
\int\frac{\dd^2q}{(2\pi)^2}\,
\tilde {\mathcal{V}}_{\rm RPA}(\ii\Omega,{\bf q})
\left[
\frac{\tilde{v}_F\sigma_i}{\Omega^2+\tilde{v}_F^2q^2}
-
\frac{2\tilde{v}_F^2q_i\bigl(\ii\Omega+\tilde{v}_F\,\boldsymbol{\sigma}\!\cdot\!{\bf q}\bigr)}
{(\Omega^2+\tilde{v}_F^2q^2)^2}
\right].
\label{eq:dSigma_dk_k0_appendix}
\end{equation}
Substituting this into Eq.~\eqref{eq:proj_k_appendix} and using
\begin{equation}
\Tr(\sigma_i\sigma_j)=2\delta_{ij},
\qquad
\Tr(\sigma_i)=0,
\label{eq:Pauli_trace_appendix}
\end{equation}
one finds
\begin{equation}
\sum_i\Tr\!\left[
\sigma_i
\left(
\frac{\tilde{v}_F\sigma_i}{\Omega^2+\tilde{v}_F^2q^2}
-
\frac{2\tilde{v}_F^2q_i(\ii\Omega+\tilde{v}_F\,\boldsymbol{\sigma}\!\cdot\!{\bf q})}
{(\Omega^2+\tilde{v}_F^2q^2)^2}
\right)
\right]
=
\frac{4\tilde{v}_F}{\Omega^2+\tilde{v}_F^2q^2}
-
\frac{4\tilde{v}_F^3q^2}{(\Omega^2+\tilde{v}_F^2q^2)^2}.
\label{eq:trace_result_spatial_appendix}
\end{equation}
Therefore
\begin{equation}
\Sigma_k
=
-\int\frac{\dd\Omega}{2\pi}
\int\frac{\dd^2q}{(2\pi)^2}\,
\tilde {\mathcal{V}}_{\rm RPA}(\ii\Omega,{\bf q})
\left[
\frac{1}{\Omega^2+\tilde{v}_F^2q^2}
-
\frac{\tilde{v}_F^2q^2}{(\Omega^2+\tilde{v}_F^2q^2)^2}
\right].
\label{eq:Sigma_k_intermediate_appendix}
\end{equation}
Combining the two terms gives the compact form
\begin{equation}
\Sigma_k
=
-\int\frac{\dd\Omega}{2\pi}
\int\frac{\dd^2q}{(2\pi)^2}\,
\tilde {\mathcal{V}}_{\rm RPA}(\ii\Omega,{\bf q})
\frac{\Omega^2}{(\Omega^2+\tilde{v}_F^2q^2)^2}.
\label{eq:Sigma_k_correct_appendix}
\end{equation}
Substituting the low-energy interaction
\begin{equation}
\tilde {\mathcal{V}}_{\rm RPA}(\ii\Omega,{\bf q})
\simeq
\frac{16}{N_f \tilde{v}_F}\,
\frac{\sqrt{\Omega^2+\tilde{v}_F^2q^2}}{q^2},
\label{eq:Vrpa_low_appendix}
\end{equation}
we obtain
\begin{equation}
\Sigma_k
=
-\frac{16}{N_f \tilde{v}_F}
\int\frac{\dd\Omega}{2\pi}
\int\frac{\dd^2q}{(2\pi)^2}\,
\frac{\Omega^2}
{q^2(\Omega^2+\tilde{v}_F^2q^2)^{3/2}}.
\label{eq:Sigma_k_afterV_appendix}
\end{equation}
Introducing $x=\Omega/(\tilde{v}_F q)$, so that $\dd\Omega=\tilde{v}_F q\,\dd x$ and $\dd^2q=2\pi q\,\dd q$, yields
\begin{equation}
\Sigma_k
=
-\frac{8}{\pi^2N_f}
\int_{\mu/\tilde{v}_F}^{\alpha/\tilde{v}_F}\frac{\dd q}{q}
\int_0^{\alpha/(\tilde{v}_Fq)}\dd x\,
\frac{x^2}{(1+x^2)^{3/2}}.
\label{eq:Sigma_k_x_appendix}
\end{equation}

To extract the leading divergence, we first evaluate the $x$ integral,
\begin{equation}
J(X)\equiv \int_0^{X}\dd x\,\frac{x^2}{(1+x^2)^{3/2}}.
\end{equation}
Using
\begin{equation}
\frac{x^2}{(1+x^2)^{3/2}}
=
\frac{1}{\sqrt{1+x^2}}
-
\frac{1}{(1+x^2)^{3/2}},
\end{equation}
we obtain
\begin{equation}
J(X)
=
\operatorname{arcsinh}X-\frac{X}{\sqrt{1+X^2}}.
\label{eq:JX_result}
\end{equation}
For $X\gg1$,
\begin{equation}
J(X)
=
\ln(2X)-1+\mathcal O(X^{-2})
\simeq \ln X+\ln2 -1.
\label{eq:JX_asymptotic}
\end{equation}
Substituting $X=\alpha/(\tilde{v}_F q)$ into Eq.~\eqref{eq:Sigma_k_x_appendix}, the remaining momentum integral can be easily performed, leading to 
\begin{equation}
\Sigma_k(\mu)
\simeq
-\frac{4}{\pi^2N_f}\ln^2\!\frac{\alpha}{\mu}-\frac{8(\ln2-1)}{\pi^2N_f}\ln\frac{\alpha}{\mu}.
\label{eq:Sigma_k_doublelog_correct}
\end{equation}
\subsection{Renormalization Group Equations of Fermi Velocity and Quasiparticle Residue}

Before deriving the renormalization group (RG) equations, let us briefly comment on the running of the effective coupling constant $\alpha$. Since $\alpha$ is not a fundamental  UV cutoff but rather the scale at which the RPA-dominated low-energy effective field theory becomes valid, one might worry whether $\alpha$ runs as the cutoff is tuned. For example, if one only considers the electron self-energy correction, the effective coupling would naively change as $\alpha\rightarrow \tilde{Z}^2\alpha$. However, this is precluded by the gauge invariance of the system. According to the Ward identity, the vertex corrections (where the interaction line connects two different fermion lines) must exactly cancel the corrections originating from the fermion self-energy (i.e., $\tilde{Z}$). Therefore, the coupling $\alpha$ remains a non-running constant when we treat the RPA corrections self-consistently.

With $\alpha$ fixed, we turn to the dressed inverse propagator, which is parameterized as
\begin{equation}
G^{-1}(\ii\omega,{\bf k})
=\tilde{Z}^{-1}\left[
-\ii\omega\bigl[1+\Sigma_\omega(\omega,k)\bigr]
+
\tilde v_F\bigl[1-\Sigma_k(\omega,k)\bigr]\,
\boldsymbol{\sigma}\!\cdot\!{\bf k}\right].
\label{eq:app_Ginv_dressed}
\end{equation}
From this expression, we can identify the running quasiparticle weight $\tilde{Z}$ and the running Fermi velocity $\tilde{v}_F$ at an energy scale $\mu$:
\begin{equation}
\tilde{Z}^{-1}(\mu)=\tilde{Z}^{-1}(\alpha)(1+\Sigma_\omega(\mu)),
\qquad
\tilde{v}_F(\mu)=\tilde{v}_F(\alpha)\,\frac{1-\Sigma_k(\mu)}{1+\Sigma_\omega(\mu)}.
\label{eq:app_Zv_defs}
\end{equation}

At this stage, we must address a subtle issue regarding the $\ln^2(\alpha/\mu)$ double logarithmic divergences obtained in our one-loop calculations. In fact, this double logarithmic divergence originates from the singular behavior of the integrands in the limit $\Omega/q\to\infty$ (or equivalently $q/\Omega \to 0$) is a IR divergence comes from the fact that Coulomb potential is unscreened at $q=0$. In this zero-momentum limit, the primary effect of the fluctuating cavity gauge field is merely to phase-rotate the fermion operator \cite{son2007quantum}. However, because such a global phase rotation constitutes a gauge redundancy, its associated singular divergences must not appear in any physical, gauge-invariant observables. 

Indeed, the quasiparticle weight $\tilde{Z}$, which is closely related to $\Sigma_\omega$, is not a gauge-invariant quantity (since the fermion field $\hat{\Psi}$ itself is not gauge invariant) and consequently suffers from this severe double logarithmic suppression. In contrast, the physical Fermi velocity $\tilde{v}_F$ is a gauge-invariant observable. From Eq.~\eqref{eq:app_Zv_defs}, the renormalized velocity is approximately $\tilde{v}_F(\mu) \approx v_F(1 - \Sigma_k - \Sigma_\omega)$. Crucially, the double logarithmic terms $\frac{4}{\pi^2 N_f}\ln^2(\alpha/\mu)$ in $\Sigma_\omega$ and $-\frac{4}{\pi^2 N_f}\ln^2(\alpha/\mu)$ in $\Sigma_k$ exactly cancel each other out in the sum. This exact cancellation leaves only a universal single logarithmic divergence, which not only provides a rigorous consistency check for our one-loop calculation but also correctly governs the linear RG flow of the Fermi velocity in the non-Fermi-liquid critical regime.
As a further consistency check, we can directly evaluate the leading divergences of $\Sigma_k$ and $\Sigma_\omega$. 
\begin{align}
\Sigma_k+\Sigma_\omega  
&= -\frac{16}{N_f } \int \frac{d\Omega d^2q}{(2\pi)^3} \frac{\tilde{v}_F^2}{(\Omega^2 + \tilde{v}_F^2 q^2)^{3/2}} 
= -\frac{8}{N_f \pi^2 } \ln \left( \frac{\alpha}{\mu} \right).
\end{align}
The result agrees with the sub-leading $\ln{\frac{\alpha}{\mu}}$ divergence in Eqs.~\eqref{eq:Sigma_k_doublelog_correct} and \eqref{eq:app_Sigma_omega_result}.

To formalize this RG flow, we differentiate Eq.~\eqref{eq:app_Zv_defs} with respect to the running scale $\mu$, which yields
\begin{gather}
    \frac{\dd   \tilde{Z}^{-1}\tilde{v}_F}{\dd \mu}= \label{eq:S133} \tilde{Z}^{-1}\tilde{v}_F\frac{\dd\Sigma_k}{\dd \mu}\\
    \label{eq:S134}
    \frac{\dd   \tilde{Z}^{-1}}{\dd \mu}=  \tilde{Z}^{-1}\frac{\dd\Sigma_\omega}{\dd \mu}
\end{gather}
The difference between Eq.~\eqref{eq:S133} and $\tilde{v}_F$ times Eq.~\eqref{eq:S134} explicitly cancels the gauge-dependent double logarithms, yielding the leading-order RG equations:
\begin{equation}
\frac{\dd \ln Z}{\dd\ell}=-\frac{8\ell}{\pi^2N_f}+O\left(\frac{1}{N_f^2}\right),
\qquad
\frac{\dd \ln \tilde v_F}{\dd \ell}
=
\frac{8}{\pi^2N_f}
+
O\!\left(\frac{1}{N_f^2}\right),
\qquad
\ell=\ln\frac{\alpha}{\mu}.
\label{eq:app_RG_eqs}
\end{equation}

By integrating these RG equations, we obtain the scaling behaviors for the quasiparticle residue and the Fermi velocity:
\begin{equation}
Z_{\rm qp}(\mu)\sim
\exp\left[\frac{4}{\pi^2 N_f}\ln^2\frac{\mu}{\alpha}\right],\qquad
\tilde v_F\sim
\exp\!\left[
\frac{8}{N_f\pi^2}\ln\!\frac{\alpha}{\mu}
\right]
\end{equation}

Before proceeding, we briefly justify the instantaneous approximation used in our Dyson--Schwinger analysis of excitonic insulator in the previous section. The effective Fermi velocity $\tilde{v}_F$ is significantly renormalized toward larger values in the infrared, bounded only by the dynamically generated excitonic gap and will not flow to infinity. Consequently, in the low-energy polarization function $\Pi(q,\omega) \sim -q^2 / \sqrt{q^2 + \omega^2/\tilde{v}_F^2}$, the frequency-dependent retardation term $\omega^2/\tilde{v}_F^2$ becomes parametrically suppressed compared to the spatial momentum $q^2$. Since the interaction is overwhelmingly dominated by its spatial component,  neglecting the explicit frequency dependence constitutes a physically consistent approximation for capturing the essential critical dynamics.

Finally, we briefly discuss the low-energy behavior of the single-particle Green's function. By terminating the RG flow at the physical energy scale $\mu=\max[\omega,v_Fk]$, we construct the full Green's function:
\begin{gather}
    G(\ii\omega,{\bf k})
=
\frac{(\omega/\alpha)^{\frac{4\ln (\omega/\alpha)}{\pi^2 N_f}}}{(-\ii\omega^{}
+
 v_F(\omega/\alpha)^{-\frac{8}{N_f\pi^2}}{\boldsymbol \sigma}\cdot{\bf k})}
\end{gather}
Here, the nonanalytic structure of the elementary excitations is explicitly manifested. The fractional power-law dependence on the cutoff in the numerator arises directly from the $\ln^2(\alpha/\mu)$ divergence of the quasiparticle weight. Since the electron operator itself is not gauge invariant, the specific coefficient of this anomalous dimension depends on the choice of gauge. Nevertheless, the non-Fermi liquid nature of the system—characterized by the vanishing of the quasiparticle pole and the singular velocity enhancement—remains a robust, physical consequence of the cavity-mediated interactions.

\section{Zeroth-Landau-Level Propagator and Polarization in a Magnetic Field}

In this section, we summarize several technical ingredients used in the strong-field analysis of the main text. We first derive the free Green's function projected onto the zeroth Landau level (ZLL) for a single Dirac valley, and then introduce the corresponding massive projected propagator employed in the Dyson--Schwinger analysis. Finally, we discuss the structure of the density-density polarization bubble in the ZLL and calculate the leading inter-Landau-level form factor, which governs the dynamical screening response.

\subsection{Single Particle Green's function in the zeroth Landau level}

We consider a single Dirac cone with chirality $(-1)^a=1$ in a perpendicular magnetic field
$\mathbf B = B \hat z$. The low-energy Hamiltonian is given by
\begin{equation}
\hat H_{B,+}
=
v_F
\begin{pmatrix}
0 & \hat \Pi_x-\ii\hat\Pi_y\\
\hat\Pi_x+\ii\hat\Pi_y & 0
\end{pmatrix},
\end{equation}
where $\hat{\bm\Pi}=\hat{\bm p}+e\hat{\bm A}$ is the kinetic momentum. For definiteness, we assume $eB>0$ and work in the Landau gauge $\bm A=(0,Bx)$.

Introducing the standard ladder operators,
\begin{equation}
\hat a=\frac{\ell_B}{\sqrt2}(\hat\Pi_x-\ii\hat\Pi_y),
\qquad
\hat a^\dagger=\frac{\ell_B}{\sqrt2}(\hat\Pi_x+\ii\hat\Pi_y),
\end{equation}
with the magnetic length $\ell_B=1/\sqrt{eB}$, we have the commutation relation $[\hat a,\hat a^\dagger]=1$. The Hamiltonian takes the off-diagonal form
\begin{equation}
\hat H_B
=
\frac{\sqrt2\,v_F}{\ell_B}
\begin{pmatrix}
0 & \hat a\\
\hat a^\dagger & 0
\end{pmatrix},
\end{equation}
yielding the well-known Landau-level spectrum
\begin{equation}
E_n=\mathrm{sgn}(n)\,\frac{v_F\sqrt{2|n|}}{\ell_B}.
\end{equation}

The zeroth Landau level lies precisely at zero energy and, given $eB>0$, is fully polarized onto the lower sublattice. Its normalized eigenstates are expressed as
\begin{equation}
\Psi_{0,k_y}(\mathbf r)
=
\frac{e^{\ii k_y y}}{\sqrt{L_y}}
\begin{pmatrix}
0\\
\phi_0(x-k_y \ell_B^2)
\end{pmatrix},
\end{equation}
where
\begin{equation}
\phi_0(x)
=
(\pi \ell_B^2)^{-1/4}
\exp\!\left(-\frac{x^2}{2\ell_B^2}\right)
\end{equation}
is the ground-state wavefunction of a one-dimensional harmonic oscillator. The projector onto this ZLL subspace is 
\begin{equation}
\mathcal P_{+}=\frac{1-\sigma_z}{2},
\end{equation}
for our chosen valley and magnetic-field orientation.

The noninteracting Matsubara Green's function projected onto the ZLL is defined as
\begin{equation}
G_{0,\rm ZLL}(\ii\omega;\mathbf r,\mathbf r')
=
\sum_{k_y}
\frac{\Psi_{0,k_y}(\mathbf r)\Psi^\dagger_{0,k_y}(\mathbf r')}{-\ii\omega}.
\label{eq:G0_ZLL_spectral_app}
\end{equation}
Its spatial dependence is determined by the kernel
\begin{equation}
I(\mathbf r,\mathbf r')
=
\sum_{k_y}
\Psi_{0,k_y}(\mathbf r)\Psi^\dagger_{0,k_y}(\mathbf r').
\end{equation}
Substituting the explicit ZLL wavefunctions, we have
\begin{equation}
\Psi_{0,k_y}(\mathbf r)\Psi^\dagger_{0,k_y}(\mathbf r')
=
\frac{e^{\ii k_y(y-y')}}{L_y}
\mathcal P_{+}\,
\phi_0(x-k_y \ell_B^2)\phi_0(x'-k_y \ell_B^2).
\end{equation}
Taking the thermodynamic limit $\sum_{k_y}\to \dfrac{L_y}{2\pi}\int \dd k_y$, the kernel becomes
\begin{equation}
I(\mathbf r,\mathbf r')
=
\mathcal P_{+}
\int_{-\infty}^{\infty}\frac{\dd k_y}{2\pi}\,
e^{\ii k_y(y-y')}
\phi_0(x-k_y \ell_B^2)\phi_0(x'-k_y \ell_B^2).
\end{equation}
Inserting the Gaussian form of $\phi_0$ gives
\begin{equation}
I(\mathbf r,\mathbf r')
=
\mathcal P_{+}\frac{1}{\sqrt{\pi}\ell_B}
\int_{-\infty}^{\infty}\frac{\dd k_y}{2\pi}
\exp\!\left[
-\frac{(x-k_y \ell_B^2)^2+(x'-k_y \ell_B^2)^2}{2\ell_B^2}
+\ii k_y(y-y')
\right].
\end{equation}

By defining $X=k_y \ell_B^2$ and utilizing the algebraic identity
\begin{equation}
(x-X)^2+(x'-X)^2
=
2\left(X-\frac{x+x'}{2}\right)^2
+\frac{(x-x')^2}{2},
\end{equation}
the integration can be performed by completing the square. This yields
\begin{equation}
I(\mathbf r,\mathbf r')
=
e^{\ii\Phi(\mathbf r,\mathbf r')}
\mathcal P_{+}\,
\frac{1}{2\pi \ell_B^2}
\exp\!\left[
-\frac{(\mathbf r-\mathbf r')^2}{4\ell_B^2}
\right],
\label{eq:I_rr_app}
\end{equation}
where
\begin{equation}
\Phi(\mathbf r,\mathbf r')
=
\frac{(x+x')(y-y')}{2\ell_B^2}
\end{equation}
is the standard Schwinger phase in the Landau gauge. Substituting Eq.~\eqref{eq:I_rr_app} into Eq.~\eqref{eq:G0_ZLL_spectral_app}, we arrive at the full real-space Green's function
\begin{equation}
G_{0,\rm ZLL}(\ii\omega;\mathbf r,\mathbf r')
=
e^{\ii\Phi(\mathbf r,\mathbf r')}
\widetilde G_{0,\rm ZLL}(\ii\omega;\mathbf r-\mathbf r'),
\end{equation}
where the translation-invariant part can be expressed as
\begin{equation}
\widetilde G_{0,\rm ZLL}(\ii\omega;\mathbf r)
=
\frac{1}{-\ii\omega}\,
\mathcal P_{+}\,
\frac{1}{2\pi \ell_B^2}
\exp\!\left(-\frac{r^2}{4\ell_B^2}\right).
\label{eq:G0_ZLL_realspace_app}
\end{equation}

Performing a Fourier transform on Eq.~\eqref{eq:G0_ZLL_realspace_app}, and adopting the convention used in the main text, we obtain the projected momentum-space Green's function:
\begin{equation}
\widetilde G_{0,\rm ZLL}(\ii\omega,\mathbf q)
=
2\,e^{-q^2\ell_B^2}\,
\frac{1}{-\ii\omega}\,
\mathcal P_{+}.
\label{eq:G0_ZLL_momentum_app}
\end{equation}

To understand how dynamical mass generation modifies the spectrum, we consider a single gapped Dirac cone in a perpendicular magnetic field,
\begin{equation}
\tilde H_{B,+}
=
v_F \left( \sigma_x \hat \Pi_x + \sigma_y \hat\Pi_y \right) + m \sigma_z,
\label{eq:H_dirac_app}
\end{equation}
where $m$ represents a dynamically generated Dirac mass. In terms of the ladder operators, this reads
\begin{equation}
\tilde H_{B,+}=
\begin{pmatrix}
m & \omega_B \hat a \\
\omega_B \hat a^\dagger & -m
\end{pmatrix},
\qquad
\omega_B = \frac{\sqrt{2}v_F}{\ell_B} = v_F \sqrt{2 e B}.
\label{eq:H_matrix_app}
\end{equation}

For $n\ge 1$, the Landau-level energies are shifted to
\begin{equation}
E_{n,\lambda} = \lambda \epsilon_n,
\qquad
\epsilon_n = \sqrt{m^2 + n \omega_B^2},
\qquad
\lambda = \pm.
\end{equation}
The corresponding eigenstates mix the sublattice components:
\begin{equation}
\Psi_{n,\lambda,k_y}(\mathbf r)
=
\frac{e^{i k_y y}}{\sqrt{L_y}}
\begin{pmatrix}
u_{n,\lambda}\,\phi_{n-1}(x-x_0) \\
v_{n,\lambda}\,\phi_n(x-x_0)
\end{pmatrix},
\qquad
x_0=-\ell_B^2 k_y,
\end{equation}
with coefficients
\begin{equation}
u_{n,\lambda}=\sqrt{\frac{\epsilon_n+\lambda m}{2\epsilon_n}},
\qquad
v_{n,\lambda}=\lambda \sqrt{\frac{\epsilon_n-\lambda m}{2\epsilon_n}}.
\end{equation}

However, the zeroth Landau level ($n=0$) remains uniquely immune to this sublattice mixing. For $B>0$, it retains the form
\begin{equation}
\Psi_{0,k_y}(\mathbf r)
=
\frac{e^{i k_y y}}{\sqrt{L_y}}
\begin{pmatrix}
0 \\
\phi_0(x-x_0)
\end{pmatrix},
\qquad
E_0=-m.
\label{eq:ZLL_state_app}
\end{equation}
Thus, the ZLL remains fully polarized on a single sublattice. In the massless limit $m\to 0$, $E_0 \to 0$, while the first excited level approaches the cyclotron scale:
\begin{equation}
\epsilon_1 = \sqrt{m^2+\omega_B^2}
\;\;\xrightarrow{m\to0}\;\;
\omega_B = v_F \sqrt{2eB}.
\end{equation}

For the interacting problem discussed in the main text, we seek a self-consistent symmetry-breaking solution where a mass gap $m(\mathbf{q})$ is dynamically generated within the ZLL sector. The dressed projected propagator used in the strong-field Dyson--Schwinger equation is therefore formulated as
\begin{equation}
\widetilde G_{\rm ZLL}(\ii\omega,\mathbf q)
=
2\,e^{-q^2\ell_B^2}\,
\frac{\ii\omega+m(q)}{\omega^2+m^2(q)}\,
\mathcal P_{+}.
\label{eq:G_ZLL_momentum_massive_app}
\end{equation}
Since the propagator is rigorously projected onto the ZLL subspace, its internal matrix structure is completely described by $\mathcal P_{+}$. The opposite Dirac valley, characterized by chirality $(-1)^a=-1$, follows analogously.

\subsection{One-loop vacuum density-density polarization in the projected ZLL}

To evaluate the one-loop density-density polarization bubble in a magnetic field, it is most convenient to compute it in real space, where the gauge-dependent Schwinger phases can be handled seamlessly. The bubble is given by
\begin{equation}
\Pi_B(x, x') =- N_f \Tr \left[ G_{\text{ZLL}}(x, x') G_{\text{ZLL}}(x', x) \right].
\end{equation}
Since the Schwinger phases from the two propagators cancel exactly, the spatial part of the bubble is simply the square of the Gaussian envelope:
\begin{equation}
f(\Delta\mathbf{r}) f(-\Delta\mathbf{r}) = \left( \frac{1}{2\pi \ell_B^2} \right)^2 \exp\left( -\frac{(\Delta\mathbf{r})^2}{2\ell_B^2} \right).
\end{equation}
Performing the two-dimensional Fourier transform with respect to the external momentum $\mathbf q$ yields
\begin{equation}
\Pi_{\text{space}}(\mathbf q)
=
\int \dd^2(\Delta\mathbf{r}) \, e^{-\ii\mathbf{q}\cdot\Delta\mathbf{r}}
\left( \frac{1}{2\pi \ell_B^2} \right)^2
e^{-\frac{(\Delta\mathbf{r})^2}{2\ell_B^2}}
=
\frac{1}{2\pi \ell_B^2} e^{-q^2 \ell_B^2 / 2}
=
\frac{eB}{2\pi} e^{-q^2 \ell_B^2 / 2}.
\end{equation}
The prefactor $eB/(2\pi)$ exactly reproduces the Landau-level degeneracy per unit area, consistent with physical expectations.

Taking the matrix trace $\Tr[\mathcal P_{+}]=1$, the full polarization bubble factorizes in Euclidean frequency-momentum space as
\begin{equation}
\Pi(\mathbf{q}, \ii\Omega)
=
-N_f \frac{eB}{2\pi} e^{-q^2 \ell_B^2 / 2}\,
\Pi_{B,\text{freq}}(\ii\Omega),
\label{eq:Pi_magnetic_app}
\end{equation}
where the frequency-dependent part takes the form of an effective $0+1$-dimensional convolution:
\begin{equation}
\Pi_{B,\text{freq}}(\ii\Omega) =
\int_{-\infty}^{\infty} \frac{\dd\omega}{2\pi}
\frac{m^2(q) - \omega(\omega+\Omega)}
{(\omega^2 + m^2(q))\left[(\omega+\Omega)^2 + m^2(q)\right]}.
\label{eq:Pi_freq_exact_app}
\end{equation}

To determine the static Thomas--Fermi screening, we consider the limit $\Omega\to 0$, which simplifies Eq.~\eqref{eq:Pi_freq_exact_app} to
\begin{equation}
\Pi_{B,\text{freq}}(0)
=
\int_{-\infty}^{\infty} \frac{\dd\omega}{2\pi}
\frac{m^2(q) - \omega^2}{(\omega^2 + m^2(q))^2}.
\end{equation}
By rewriting the numerator as $m^2-\omega^2=(m^2+\omega^2)-2\omega^2$, we can partition the integral into two pieces:
\begin{equation}
\Pi_{B,\text{freq}}(0)
=
\int_{-\infty}^{\infty} \frac{\dd\omega}{2\pi}\frac{1}{\omega^2+m^2(q)}
-
\int_{-\infty}^{\infty} \frac{\dd\omega}{2\pi}\frac{2\omega^2}{(\omega^2+m^2(q))^2}.
\end{equation}
The first integral gives
\begin{equation}
I_1
=
\int_{-\infty}^{\infty} \frac{\dd\omega}{2\pi}\frac{1}{\omega^2+m^2(q)}
=
\frac{1}{2m(q)},
\end{equation}
while the second evaluates to the same value via integration by parts:
\begin{equation}
I_2
=
\int_{-\infty}^{\infty} \frac{\dd\omega}{2\pi}\frac{2\omega^2}{(\omega^2+m^2(q))^2}
=
\frac{1}{2m(q)}.
\end{equation}
Consequently, we find
\begin{equation}
\Pi_{B,\text{freq}}(0)=I_1-I_2=0.
\end{equation}
The static density response vanishes identically within the projected ZLL. Physically, this reflects the incompressibility of the completely filled (or empty) states, dictating the absence of a regular intra-ZLL density response in the gapped projected theory.

Because the intra-ZLL channel is suppressed, for finite frequencies $\Omega\neq 0$, higher Landau levels must be included to capture the physical absorption. The leading nonvanishing dynamical contribution in the strong-field regime originates from inter-level transitions between the zeroth and first Landau levels ($n=0 \leftrightarrow 1$). Let us now evaluate the corresponding form factor for massless Dirac fermions.

The density operator is given by $\rho(\mathbf q)=e^{i\mathbf q\cdot \mathbf r}$. At zero temperature, the density-density polarization function admits the standard spectral representation
\begin{equation}
\Pi_B(\ii\omega,\mathbf q)
=
\frac{1}{S}
\sum_{a,b}
\frac{f(E_a)-f(E_b)}
{-\ii\omega + E_a - E_b }
\,
\left|
\langle a|\hat\rho(\mathbf q)|b\rangle
\right|^2,
\label{eq:polarization_spectral_app}
\end{equation}
where $(2\pi \ell_B^2)^{-1}$ is the density of states per Landau level, and the indices $a,b$ run over all exact Landau-level quantum numbers including band indices.

In the Landau level basis, we have
\begin{gather}
    \hat\rho(\vq)=\sum_{\lambda,\lambda'}\sum_{n,n'}\sum_{k,k'}\langle n',k',\lambda' | e^{i\mathbf q\cdot \mathbf r} | n,k,\lambda \rangle \hat{c}^\dagger_{n',k',\lambda'}\hat{c}_{n,k,\lambda},
\end{gather}
and we only keep with the contribution of zeroth and first Landau levels, leading to
\begin{gather}
    \hat\rho(\vq)\sim\hat{\rho}_0(\vq)+\sum_{\lambda}\sum_{k,k'}\langle 0,k' | e^{i\mathbf q\cdot \mathbf r} | 1,k,\lambda \rangle \hat{c}^\dagger_{0,k'}\hat{c}_{1,k,\lambda}+{\rm H.c.},
\end{gather}
where $\hat{\rho}_0(\vq)$ is the ZLL projecting density operator.
For a neutral system, the dominant low-energy transition providing the lowest excitation energy allowed by the density operator is the $n=0 \leftrightarrow 1$ channel. The essential matrix element required is
\begin{equation}
F_{01}(\mathbf q)
=
\langle 0,k' | e^{i\mathbf q\cdot \mathbf r} | 1,k \rangle,
\end{equation}
where $\lambda$ is dropped off for convenience, since it only gives an overall $\pm$ factor.

The single-particle Landau-level wavefunctions in the Landau gauge take the form
\begin{equation}
\psi_{n,k}(x,y)
=
\frac{1}{\sqrt{L_y}}\,e^{iky}\,\varphi_n(x-x_k),
\qquad
x_k\equiv k\ell_B^2,
\end{equation}
where $\varphi_n(\xi)$ are harmonic-oscillator eigenfunctions. Specifically, we need the first two states:
\begin{align}
\varphi_0(\xi)&=\frac{1}{\pi^{1/4}\sqrt{\ell_B}}\,e^{-\xi^2/(2\ell_B^2)},\\
\varphi_1(\xi)&=\sqrt{2}\,\frac{\xi}{\ell_B}\,\varphi_0(\xi).
\end{align}
We define the transition integral explicitly as
\begin{align}
F_{01}(\mathbf q;k',k)
&\equiv \langle 0,k'|e^{i\mathbf q\cdot\mathbf r}|1,k\rangle \notag\\
&=
v_{1,\lambda}\int d^2r\;\psi^*_{0,k'}(x,y)\,e^{i(q_x x+q_y y)}\,\psi_{1,k}(x,y).
\end{align}

Integrating over the $y$ coordinate yields the momentum conservation condition:
\begin{align}
\frac{1}{L_y} \int_0^{L_y} dy\; e^{i(k-k'+q_y)y} = \delta_{k',\,k+q_y}.
\end{align}
This allows us to write $F_{01}(\mathbf q;k',k) = \delta_{k',\,k+q_y}\; I_{01}(q_x;k)$, where the residual integration over $x$ is
\begin{equation}
I_{01}(q_x;k)
=
\int dx\;\varphi_0(x-x_{k'})\,e^{iq_x x}\,\varphi_1(x-x_k),
\qquad (k'=k+q_y).
\end{equation}
Since the shift between the center coordinates is $x_{k'}-x_k = (k'-k)\ell_B^2 = q_y\ell_B^2 \equiv \delta$, we can evaluate this analytically. By shifting the integration variable to $u=x-x_k$ (and defining $a\equiv x_k$), we get
\begin{align}
I_{01}
&=\int du\;\varphi_0(u-\delta)\,e^{iq_x(u+a)}\,\varphi_1(u)\notag\\
&=
e^{iq_x a}\,\frac{\sqrt{2}}{\ell_B}
\int du\;u\;\varphi_0(u-\delta)\varphi_0(u)\,e^{iq_x u}.
\end{align}
The product of the Gaussian functions simplifies to
\begin{align}
\varphi_0(u-\delta)\varphi_0(u)
&=
\frac{1}{\sqrt{\pi}\,\ell_B}
\exp\!\left[-\frac{(u-\delta)^2+u^2}{2\ell_B^2}\right]\notag\\
&=
\frac{1}{\sqrt{\pi}\,\ell_B}
\exp\!\left[-\frac{(u-\delta/2)^2}{\ell_B^2}\right]
\exp\!\left[-\frac{\delta^2}{4\ell_B^2}\right].
\end{align}
Inserting this back, we obtain
\begin{align}
I_{01}
=
e^{iq_x a}\,\frac{\sqrt{2}}{\ell_B}\,
\frac{e^{-\delta^2/(4\ell_B^2)}}{\sqrt{\pi}\,\ell_B}
\int du\;u\,e^{-(u-\delta/2)^2/\ell_B^2}\,e^{iq_x u}.
\end{align}
Making a second shift $t=u-\delta/2$, the integral becomes
\begin{equation}
\int du\;u\,e^{-(u-\delta/2)^2/\ell_B^2}\,e^{iq_x u}
=
e^{iq_x\delta/2}\int dt\;(t+\delta/2)\,e^{-t^2/\ell_B^2}\,e^{iq_x t}.
\end{equation}
Using the standard Gaussian integrals,
\begin{align}
\int dt\;e^{-t^2/\ell_B^2}e^{iq_x t}
&=
\sqrt{\pi}\,\ell_B\,e^{-q_x^2\ell_B^2/4},\\
\int dt\;t\,e^{-t^2/\ell_B^2}e^{iq_x t}
&=
i\frac{\sqrt{\pi}\,\ell_B^3}{2}q_x\,e^{-q_x^2\ell_B^2/4},
\end{align}
and assembling all terms, we find
\begin{align}
I_{01}
&=
e^{iq_x(a+\delta/2)}\,
e^{-\delta^2/(4\ell_B^2)}\,e^{-q_x^2\ell_B^2/4}\;
\frac{\sqrt{2}}{\ell_B}
\left(\frac{\delta}{2}+ i\frac{\ell_B^2 q_x}{2}\right)\notag\\
&=
e^{iq_x(x_k+x_{k'})/2}\;
\exp\!\left[-\frac{\ell_B^2(q_x^2+q_y^2)}{4}\right]\;
\frac{\ell_B}{\sqrt{2}}\,(q_y+i q_x),
\end{align}
where we have reinstated $\delta=q_y\ell_B^2$.

Combining these ingredients, we arrive at the exact matrix element:
\begin{equation}
\langle 0,k'|e^{i\mathbf q\cdot\mathbf r}|1,k\rangle
=v_{1,\lambda}
\delta_{k',\,k+q_y}\;
e^{\,i q_x (x_k+x_{k'})/2}\;
\frac{\ell_B}{\sqrt{2}}\,(q_y+i q_x)\;
\exp\!\left(-\frac{q^2\ell_B^2}{4}\right)
\label{eq:Ffactor}
\end{equation}
with $q^2=q_x^2+q_y^2$. In most physical applications, only its modulus squared enters the calculation, in which case the phase factor smoothly drops out:
\begin{equation}
\left|\langle 0,k'|e^{i\mathbf q\cdot\mathbf r}|1,k\rangle\right|^2
=|v_{1,\lambda}|^2
\delta_{k',\,k+q_y}\;
\frac{\ell_B^2 q^2}{2}\;
\exp\!\left(-\frac{q^2\ell_B^2}{2}\right).
\end{equation}
We then sum over $k_y,k_y'$
\begin{equation}
\frac{1}{S}\sum_{k_y,k_y'}
\left|F_{01}(\mathbf q;k_y',k_y)\right|^2
=
\frac{|v_{1,\lambda}|^2}{4\pi}
q^2 e^{-q^2\ell_B^2/2},
\label{eq:ky_summed_form_factor_section}
\end{equation}
where we use $\sum_{k_y}1
=
\frac{S}{2\pi \ell_B^2}.$ This form factor controls the leading inter-Landau-level contribution to the dynamical density response in the strong-field regime.

Substituting Eq.~\eqref{eq:Ffactor} into Eq.~\eqref{eq:polarization_spectral_app}, the polarization function evaluates to
\begin{align}
\Pi_B(\ii\omega,\mathbf q)
&=
\sum_{\lambda}
\frac{|v_{1,\lambda}|^2}{4\pi}
q^2 e^{-q^2\ell_B^2/2}
\left[
\frac{f(E_0)-f(E_{1,\lambda})}
{-\ii\omega+E_0-E_{1,\lambda}}
+
\frac{f(E_{1,\lambda})-f(E_0)}
{-\ii\omega+E_{1,\lambda}-E_0}
\right].
\label{eq:Pi01_after_ky_sum}
\end{align}
At zero temperature and exact charge neutrality, only appropriate interband transitions contribute. For the massless Dirac case, $E_0=0$ and $E_{1,\lambda}=\lambda\omega_B$. The two terms combine into a standard resonant denominator:
\begin{equation}
\Pi_B(\ii\omega,\mathbf q)
=\frac{N_f}{8\pi}
q^2 e^{-q^2\ell_B^2/2}
\left[
\frac{1}{\ii\omega-\omega_B}
-
\frac{1}{\ii\omega+\omega_B}
\right].
\end{equation}
Combining these fractions yields the dynamical polarization:
\begin{equation}
\Pi_B(\ii\omega,\mathbf q)
=-\frac{N_f}{8\pi}
q^2 e^{-q^2\ell_B^2/2}
\frac{2\omega_B}{\omega^2+\omega_B^2}.
\end{equation}

We can now revisit the gap equation in the strong-field regime. Taking the bare interaction kernel to be
\begin{equation}
\mathcal{V}_0(\mathbf q)=-\frac{\alpha}{q^2},
\end{equation}
and assuming that the magnetic field is sufficiently strong, the dominant screening effect can be described within the quasistatic approximation. In this regime, the dynamical polarization from the $n=0 \leftrightarrow 1$ transition reduces to its static limit,
\begin{equation}
\Pi_B(0,\mathbf q)=-\frac{N_f}{4\pi\omega_B}q^2 e^{-q^2\ell_B^2/2}.
\end{equation}

The fully screened effective interaction is therefore given by
\begin{equation}
V_{\rm eff}(\mathbf q)
=
\frac{1}{\mathcal{V}_0^{-1}(\mathbf q)-\Pi(0,\mathbf q)}
=
\frac{1}{-q^2/\alpha+\alpha_B q^2 e^{-q^2 \ell_B^2/2}}
=
\frac{1}{q^2}\,
\frac{1}{-\alpha^{-1}+\alpha_B e^{-q^2 \ell_B^2/2}}.
\label{eq:Veff_static}
\end{equation}

In the long-wavelength limit $q\ell_B \ll 1$, the dressed interaction is given by 
\begin{equation}
    \mathcal{V}(q) = \frac{\alpha_{\text{eff}}}{q^2},
\end{equation}
where the effective coupling strength $\alpha_{\text{eff}}$ satisfies
\begin{equation}
    \alpha_{\text{eff}}^{-1} =  \frac{N_f}{4\pi\omega_B} - \alpha^{-1} .
\end{equation}
The system exhibits a critical magnetic field that marks a transition from the \textit{underscreening} to the \textit{overscreening} regime. 

Specifically, when $\omega_B \ll \alpha$, the screening effect dominates, and the effective interaction takes the form
\begin{equation}
    \mathcal{V}(q) = \frac{4\pi\omega_B}{N_f} \frac{1}{q^2} e^{q^2\ell_B^2/2}.
\end{equation}
Note that the exponential factor $e^{q^2\ell_B^2/2}$ does not lead to a physical divergence; in the Landau level basis, the correlation functions are naturally regularized by the form factor $e^{-q^2\ell_B^2/2}$, ensuring the stability of the theory.

In the opposite limit where $\omega_B \gg \alpha$, the interaction enters the overscreening regime and becomes purely attractive:
\begin{equation}
    \mathcal{V}(q) = -\frac{\alpha}{q^2}.
\end{equation}

\subsection{Dyson--Schwinger equation and mass generation in the ZLL}

In this section, we provide the detailed derivation of the gap equation in the strong magnetic field limit, as discussed in the main text. Starting from the projected Dyson--Schwinger equation,
\begin{equation}
\widetilde G_{\rm ZLL}^{-1}(\ii\omega;\mathbf r)
=
\widetilde G_{0,\rm ZLL}^{-1}(\ii\omega;\mathbf r)
+
\int\frac{\dd\Omega}{2\pi}
\int\frac{\dd^2q}{(2\pi)^2}\,
\mathcal{V}(\mathbf q)\,
e^{\ii\mathbf q\cdot\mathbf r}\,
\widetilde G_{\rm ZLL}(\ii\omega+\ii\Omega;\mathbf r),
\label{eq:Dyson_ZLL_appendix}
\end{equation}
we seek a self-consistent symmetry-breaking solution in which the cavity-mediated interactions dynamically generate a mass gap $m$ entirely within the ZLL.

For the Dirac cone with chirality $(-1)^a=1$, we employ the massive projected propagator ansatz:
\begin{equation}
\widetilde G_{\rm ZLL}(\ii\omega,\mathbf q)
=
2\,e^{-q^2\ell_B^2}\,
\frac{\ii\omega+m(q)}{\omega^2+m^2(q)}\,
\mathcal P_{+}.
\label{eq:G_ZLL_ansatz_appendix}
\end{equation}
Similarly, for the opposite Dirac valley with chirality $(-1)^a=-1$, the corresponding ansatz takes the form:
\begin{equation}
\widetilde G_{\rm ZLL}(\ii\omega,\mathbf q)
=
2\,e^{-q^2\ell_B^2}\,
\frac{\ii\omega-m(q)}{\omega^2+m^2(q)}\,
\mathcal P_{-}.
\label{eq:G_ZLL_ansatz_minus_appendix}
\end{equation}

Assuming the bare cavity-induced interaction kernel,
\begin{equation}
\mathcal{V}(\mathbf q)\simeq \frac{\alpha_{eff}}{q^2},
\label{eq:bare_int_appendix}
\end{equation}
one can project Eq.~\eqref{eq:Dyson_ZLL_appendix} onto the ZLL mass channel. Since the full propagator is rigorously restricted to the ZLL subspace, its matrix structure is entirely encoded by the projection operator $\mathcal P_{\pm}$. Consequently, the Dyson--Schwinger equation simplifies into a scalar integral equation for the dynamically generated gap $m(q)$:
\begin{equation}
\frac{m(k)}{2}
=
\int\frac{\dd\Omega}{2\pi}
\int\frac{\dd^2q}{(2\pi)^2}\,
\mathcal{V}(\mathbf q)\,
e^{-(q^2+2qk\cos\theta)\ell_B^2}\,
\frac{2m(q)}{\Omega^2+m^2(q)}.
\label{eq:gap_start_appendix}
\end{equation}
When $k=0$, we have
\begin{equation}
m(0)
=
4\int\frac{\dd\Omega}{2\pi}
\int\frac{\dd^2q}{(2\pi)^2}\,
\mathcal{V}(\mathbf q)\,
e^{-q^2\ell_B^2}\,
\frac{m(q)}{\Omega^2+m^2(q)}.
\label{eq:gap_start_appendix}
\end{equation}

A remarkable and defining feature of dynamical mass generation within the zeroth Landau level is the effective dimensional reduction from $2+1$ to $0+1$ dimensions. Specifically, the temporal and spatial loop integrations factorize completely. Performing the Matsubara frequency integral yields exactly:
\begin{equation}
\int\frac{\dd\ \Omega}{2\pi}\,
\frac{m(q)}{\Omega^2+m^2(q)}
=
\frac{1}{2},
\qquad (m>0).
\label{eq:freq_integral_appendix}
\end{equation}
Because this integral is simply a constant, the mass function $m(q)$ miraculously drops out of the remaining momentum integration. As a result, the gap equation determines a uniform mass scale and reduces to:
\begin{equation}
m(0)
=
2\int\frac{\dd^2q}{(2\pi)^2}\,
\mathcal{V}(\mathbf q)\,
e^{-q^2\ell_B^2}.
\label{eq:gap_momentum_appendix}
\end{equation}
Substituting the cavity interaction Eq.~\eqref{eq:bare_int_appendix} and transitioning to polar coordinates ($\dd^2q=2\pi q\,\dd q$). The integrand exhibits a $1/q$ behavior at small momenta, signaling a severe logarithmic infrared singularity. However, within the physical metasurface cavity platform, the idealized bare form $\mathcal{V}(\mathbf q)\simeq \alpha/q^2$ derived from the quasistatic approximation is no longer valid at length scales exceeding the cavity wavelength. This divergence is naturally cut off at a characteristic infrared momentum scale $k_p \sim \omega_p/c$, where $\omega_p$ is the effective cavity plasma frequency. By implementing this physical IR cutoff, the regularized gap equation becomes
\begin{equation}
m(0)
=
\frac{4\omega_B}{N_f}
\int_{k_p}^{\infty}\frac{\dd q}{q}\,
e^{-q^2\ell_B^2/2}.
\label{eq:gap_reg_appendix}
\end{equation}
where $q_{\rm UV}\sim 2\pi/d$, beyond which the electron cannot feel the existence of metasurfaces.

This integral can be evaluated analytically in terms of the exponential integral function. With the change of variable
\begin{equation}
u=\frac{q^2\ell_B^2}{2}
\qquad
\frac{\dd q}{q}=\frac{\dd u}{2u},
\end{equation}
Eq.~\eqref{eq:gap_reg_appendix} transforms into
\begin{equation}
m(0)
=
\frac{2\omega_B}{N_f}
\int_{k_p^2 \ell_B^2/2}^{\infty}
\frac{e^{-u}}{u}\,\dd u
=
\frac{\omega_B}{2N_f}
E_1\!\left({k_p^2 \ell_B^2/2}\right),
\label{eq:gap_E1_appendix}
\end{equation}
where $E_1(x)$ is the standard exponential integral,
\begin{equation}
E_1(x)=\int_x^\infty \frac{e^{-u}}{u}\,\dd u.
\end{equation}
In the strong-field regime where $k_p \ell_B \ll 1$, we can expand $E_1(x)$ using its small-argument asymptotic form,
\begin{equation}
E_1(x)=-\gamma-\ln x+\mathcal O(x),
\end{equation}
where $\gamma \approx 0.5772$ is the Euler--Mascheroni constant. Substituting this expansion gives
\begin{align}
m(0)
&\simeq
\frac{2\omega_B}{N_f}
\left[
-\gamma-\ln\!\left({k_p^2\ell_B^2/2}\right)
\right]
\nonumber\\
&=
\frac{2\omega_B}{N_f}
\ln\!\left(
\frac{\,\sqrt{2}e^{-\gamma/2}}{k_p \ell_B}
\right).
\end{align}
Finally, using the definition of the magnetic length $\ell_B=1/\sqrt{e B}$,  we arrive at the expression for the dynamically generated magnetic mass:
\begin{equation}
m(0)
\simeq
\frac{2\omega_B}{N_f}
\ln\!\left(
\frac{2\text{e}^{-\gamma}eB}{\hbar k_p^2}
\right),
\end{equation}
which is in precise agreement with the scaling behavior quoted in the main text. The approximation neglecting transitions to higher Landau levels is self-consistent provided that $m(0) \ll \omega_B$. Consequently, this result remains reliable in the large-$N_f$ limit.

\subsection{Interlayer pairing in a magnetic field}

In this section, we formulate the interlayer pairing instability for two neutral Dirac layers in a perpendicular magnetic field. We consider two graphene-like layers, labeled by $s=\uparrow,\downarrow$, and restrict ourselves to a single valley. Both layers are fixed at charge neutrality throughout this section. We work in the Landau-level basis and project onto the ZLL, where the single-particle kinetic energy is quenched and the low-energy physics is governed by the projected interaction.

We denote the ZLL annihilation operator in layer $s$ by $c_{s,k}(\tau)$, where $k$ is the momentum quantum number. In the projected ZLL subspace, we introduce the standard two-component Nambu spinor in imaginary time,
\begin{equation}
\hat\Psi_k = 
\begin{pmatrix}
\hat c_{\uparrow,k}\\
\hat c^\dagger_{\downarrow,-k}
\end{pmatrix}.
\label{eq:Nambu_time_domain_clean}
\end{equation}
We use $\sigma_i$ to denote Pauli matrices acting on the layer space $(\uparrow,\downarrow)$. In the absence of pairing, the inverse Green's function in Matsubara space is
\begin{equation}
\mathcal{G}_0^{-1}(\ii\omega;k) = -\frac{\ii}{2}\omega e^{k^2\ell_B^2}\,\sigma_0,
\label{eq:G0inv_Nambu_clean}
\end{equation}

We now consider the interlayer Cooper channel. Since the two fermions are exchanged between the two layers, the pairing matrix in the layer space must be antisymmetric and is therefore proportional to $\ii\sigma_2$.  By choosing a gauge where the pairing amplitude is purely real, interlayer pairing matrix in the ZLL basis is then
\begin{equation}
\bm{\Delta}_{k} = \frac{1}{2}\Delta(k)\, \sigma_1,
\label{eq:Deltahat_matrix_clean}
\end{equation}
. In the standard Nambu basis \eqref{eq:Nambu_time_domain_clean}, the BdG self-energy explicitly incorporates the Landau-level form factor and takes the form
\begin{equation}
\Sigma_{\rm BdG}(k) = \frac{e^{k^2\ell_B^2}}{2} \Delta(k)\, \sigma_1
\end{equation}

The full inverse Green's function is obtained from the Dyson equation
\begin{equation}
\mathcal{G}^{-1}(i\omega;k) = \mathcal{G}_0^{-1}(i\omega;k) - \Sigma_{\rm BdG}(k).
\label{eq:Dyson_inverse_clean}
\end{equation}
For the uniform ansatz this gives
\begin{equation}
\mathcal{G}^{-1}(i\omega;k) = \frac{1}{2}\left[ -i\omega\sigma_0 + \Delta(k)\,\sigma_1 \right] e^{k^2\ell_B^2}.
\label{eq:Ginv_uniform_clean}
\end{equation}
 we obtain
\begin{equation}
\mathcal{G}(i\omega;k) = 2{e^{-k^2\ell_B^2}} \frac{ i\omega\,\sigma_0 + \Delta(k)\,\sigma_1 }{ \omega^2+\Delta(k)^2 }.
\label{eq:Gfull_uniform_clean}
\end{equation}
The anomalous block (upper right element) of the Green's function is therefore
\begin{equation}
\mathcal{F}(i\omega;k) = 2e^{-k^2\ell_B^2} \frac{\Delta(k)\,}{\omega^2+\Delta(k)^2}.
\label{eq:anomalous_block_clean}
\end{equation}

We now include the cavity-mediated interlayer attraction. In momentum space, we take
\begin{equation}
\mathcal{V}(q)=-\frac{\alpha}{q^2}, \qquad \alpha>0.
\label{eq:bare_V_clean}
\end{equation}
And it contribute an additional $\sigma_3$ when interacting with the vertex,The BCS channel Dyson-Schwingle equation gives
\begin{equation}
\frac{{\Delta}(0)}{2} = - \int\frac{\dd\omega}{2\pi} \int\frac{d^2q}{(2\pi)^2}\, \mathcal{V}(q)\, \mathcal{F}(i\omega;q).
\label{eq:matrix_gap_equation_clean}
\end{equation}
Substituting Eq.~\eqref{eq:anomalous_block_clean}, we obtain
\begin{equation}
\Delta(0)= - 4\int\frac{\dd\omega}{2\pi} \int\frac{d^2q}{(2\pi)^2}\, \left(-\frac{\alpha}{q^2}\right) \left(e^{-q^2\ell_B^2}\frac{\Delta(q)\,}{\omega^2+\Delta(q)^2}\right).
\label{eq:matrix_gap_equation_sub_clean}
\end{equation}
Since both sides are proportional to the same antisymmetric tensor $i\sigma_2$, this reduces to a scalar equation for $\Delta(q)$:
\begin{equation}
\Delta(0) = 4\alpha \int\frac{d^2q}{(2\pi)^2} \frac{e^{-q^2\ell_B^2}}{q^2} \int\frac{\dd\omega}{2\pi} \frac{\Delta(q)}{\omega^2+\Delta(q)^2}.
\label{eq:scalar_gap_equation_explicit_clean}
\end{equation}
At zero temperature, the frequency integral is
\begin{equation}
\int\frac{\dd\omega}{2\pi} \frac{\Delta(q)}{\omega^2+\Delta(q)^2} = \frac{\Delta(q)}{2|\Delta(q)|}.
\label{eq:omega_integral_clean}
\end{equation}
Hence, for a nontrivial solution $\Delta(q)>0$, the self-consistency equation gives
\begin{equation}
\Delta(0)
=
\frac{\alpha}{2 \pi}
\ln\!\left(
\frac{e^{-\gamma}}{k_p^2\ell_B^2}
\right).
\label{eq:Delta_gap_final}
\end{equation}

\section{Detailed estimation of experimental parameters}

Here we give further details about the numerical estimates underlying the discussion in the main text. 
As a representative metasurface platform, we consider the graphene nanodisk arrays reported in Ref.~{\cite{fang2014active}}, where the disk diameter is $D\approx 60~\mathrm{nm}$ and the edge-to-edge gap is $\delta\approx 30~\mathrm{nm}$. 
The corresponding lattice period is therefore
\begin{equation}
    a \approx D+\delta \approx 90~\mathrm{nm}.
\end{equation}

For the experimentally relevant doping range, the measured plasmonic resonance of the array lies near $\hbar\omega_0\approx 0.17$--$0.18~\mathrm{eV}$. 
This corresponds to a free-space wavelength
\begin{equation}
\lambda_0=\frac{hc}{\hbar\omega_0}\approx 6.9\text{--}7.3~\mu\mathrm{m},
\end{equation}
so that
\begin{equation}
\frac{a}{\lambda_0}\approx \frac{90~\mathrm{nm}}{7~\mu\mathrm{m}}
\approx 1.3\times 10^{-2}\ll 1.
\end{equation}
Thus the effective-medium description of the patterned graphene layer as a homogenized high-impedance surface is well justified.

To obtain a concrete order-of-magnitude estimate for the effective surface impedance, we take a representative unit-cell capacitance
\begin{equation}
C_s=5\times 10^{-18}~\mathrm{F}.
\end{equation}
Using $\hbar\omega_0\approx 0.18~\mathrm{eV}$, the corresponding angular frequency is
\begin{equation}
\omega_0=\frac{0.18\times 1.602\times 10^{-19}~\mathrm{J}}{1.055\times 10^{-34}~\mathrm{J\,s}}
\approx 2.73\times 10^{14}~\mathrm{s}^{-1}.
\end{equation}
The effective inductance then follows from $\omega_0=(L_sC_s)^{-1/2}$:
\begin{equation}
L_s=\frac{1}{\omega_0^2 C_s}
\approx \frac{1}{(2.73\times 10^{14})^2 \times 5\times 10^{-18}}
\approx 2.7\times 10^{-12}~\mathrm{H}.
\end{equation}

Next we take the cavity spacing to be one tenth of the free-space wavelength,
\begin{equation}
d\sim \lambda_0/10\approx 0.7~\mu\mathrm{m},
\end{equation}
and use $\varepsilon_d=1$ for a vacuum spacer. 
The cavity mode energy is controlled by
\begin{equation}
\omega_p^2=
\frac{\omega_0^2}{1+\dfrac{\varepsilon_0\varepsilon_d d}{2C_s}}.
\end{equation}
With $\varepsilon_0=8.85\times 10^{-12}~\mathrm{F/m}$ and $d=0.7\times 10^{-6}~\mathrm{m}$, one finds
\begin{equation}
\frac{\varepsilon_0\varepsilon_d d}{2C_s}
=
\frac{(8.85\times 10^{-12})(0.7\times 10^{-6})}{2\times 5\times 10^{-18}}
\approx 0.62.
\end{equation}
Hence
\begin{equation}
\omega_p\approx \frac{\omega_0}{\sqrt{1.62}}\approx 0.79\,\omega_0,
\end{equation}
which gives
\begin{equation}
\hbar\omega_p \approx 0.14~\mathrm{eV}.
\end{equation}

The interaction scale is
\begin{equation}
\alpha=\frac{4e^2}{2C_s+\varepsilon_0\varepsilon_d d}.
\end{equation}
Substituting $e=1.602\times 10^{-19}~\mathrm{C}$, $C_s=5\times 10^{-18}~\mathrm{F}$, and $d=0.7\times 10^{-6}~\mathrm{m}$, we obtain
\begin{equation}
2C_s+\varepsilon_0\varepsilon_d d
\approx 1.0\times 10^{-17}+6.2\times 10^{-18}
=1.62\times 10^{-17}~\mathrm{F},  
\end{equation}
and therefore
\begin{equation}
\alpha
\approx
\frac{4\times(1.602\times 10^{-19})^2}{1.62\times 10^{-17}}
\approx 6.34\times 10^{-21}~\mathrm{J}
\approx 3.96\times 10^{-2}~\mathrm{eV}.  
\end{equation}

Here we give a representative estimate of the zero-field excitonic gap using the improved Miransky form
\begin{equation}
M(0)=
\alpha\,
\exp\!\left[
-\frac{2\pi-\delta_{N_f}}{\sqrt{N_c/N_f-1}}
\right],
\qquad
\delta_{N_f}=2\arctan\!\bigl(\sqrt{N_c/N_f-1}\bigr),
\label{eq:M0_improved_SM}
\end{equation}
with
\begin{equation}
\alpha=\frac{4e^2}{2C_s+\varepsilon_0\varepsilon_d d}.
\label{eq:alpha_SM}
\end{equation}
For $N_f=2$, using $N_c=16/\pi\simeq 5.09$, 
we find 
\begin{equation}
\sqrt{N_c/N_f-1}
=
\sqrt{5.09/2-1}
\simeq 1.24.
\end{equation}
This gives
\begin{equation}
\delta_{N_f}
=
2\arctan(1.24)
\simeq 1.79.
\end{equation}
The exponential factor in Eq.~(\ref{eq:M0_improved_SM}) then becomes
\begin{equation}
\exp\!\left[
-\frac{2\pi-\delta_{N_f}}{\sqrt{N_c/N_f-1}}
\right]
\simeq
\exp\!\left(-\frac{6.283-1.79}{1.24}\right)
\simeq 2.7\times10^{-2}.
\end{equation}
Substituting this together with the representative metasurface parameters adopted in the main text,
$C_s=5\times10^{-18}~\mathrm{F}$, $\varepsilon_d=1$, and
$d\simeq 0.7~\mu\mathrm{m}$,  we obtain
\begin{equation}
M(0)
\simeq
(3.96\times10^{-2}~\mathrm{eV})(2.7\times10^{-2})
\simeq
1.1~\mathrm{meV}
\end{equation}
In temperature units, this corresponds to
\begin{equation}
\frac{M(0)}{k_B}
\simeq
\frac{1.1\times10^{-3}~\mathrm{eV}}{8.617\times10^{-5}~\mathrm{eV/K}}
\simeq 12.8~\mathrm{K}.
\end{equation}
Thus, for the representative parameter set used here, the improved gap formula yields a zero-field excitonic gap of order
$M(0)\sim 10^{-3}~\mathrm{eV}$ for $N_f=2$, corresponding to a temperature scale of order $10~\mathrm{K}$.

We briefly comment on the length scale of screening, the bare $\mathcal{V}_0(r)\sim\alpha\ln r$ attractive interaction manifests only when $r$ is not significantly smaller than $d$; otherwise, the electrons remain insensitive to the surface and experience the standard $1/r$ Coulombic repulsion. However, based on the calculation of $\alpha/(\hbar v_F)$, the characteristic screening length for a Fermi velocity $v_F = 1 \times 10^6 \, \text{m/s}$ is $l_0 \approx 16 \ \text{nm}\ll d$. Consequently, the actual screening prevents electrons from perceiving the attractive interaction at any distance, eliminating concerns regarding pairing instabilities—aside from exciton pairing—driven by such attraction.

Finally, we estimate the mass term $\Delta(0)$ for the ZLL at $B = 10\text{T}$, for bilayer $N_f=2$ Dirac semimetal with $v_F=1\times10^6{\rm ms^{-1}}$, the $K/K'$ point of different layers of Dirac semimetal can paring together, we have $\omega_B\approx115 {\rm meV}$ and $\frac{4\pi\omega_B}{2N_f\alpha}\sim 10$, so the scrrening effect can be neglected. According to the gap equation, $
\Delta(0)
=
\frac{\alpha}{2 \pi}
\ln\!\left(
\frac{e^{-\gamma}}{k_p^2\ell_B^2}
\right).$. Using the experimental parameters, the magnetic length is $l_B = \sqrt{\hbar/eB} \approx 8.11\text{ nm}$. The cavity cutoff wavevector is $k_p = \omega_p / c \approx 7.1 \times 10^{-4}\text{ nm}^{-1}$ (with $\hbar\omega_p = 0.14\text{ eV}$). Substituting the interaction scale $\alpha \approx 40\text{ meV}$, we obtain \begin{gather} \Delta(0)=62\ {\rm meV} \end{gather}.

\end{document}